\documentclass[12pt]{article}
\usepackage{graphicx}
\usepackage{color}
\newcommand{\beq}{\begin{equation}}
\newcommand{\eeq}{\end{equation}}
\newcommand{\bea}{\begin{eqnarray}}
\newcommand{\eea}{\end{eqnarray}}

\newcommand{\gsim}{\lower.7ex\hbox{$
\;\stackrel{\textstyle>}{\sim}\;$}}
\newcommand{\lsim}{\lower.7ex\hbox{$
\;\stackrel{\textstyle<}{\sim}\;$}}

\newcommand{\eod}{\end{document}}

\definecolor{verm}{rgb}{0.8,0.1,0.0}

\begin{document}
\thispagestyle{empty}
\vspace*{-22mm}

\begin{flushright}
UND-HEP-12-BIG\hspace*{.08em}03\\
\today\\

%hep-ph/0703132\\

\end{flushright}
\vspace*{1.3mm}

\begin{center}
{\Large {\bf Second Generation of `Miranda Procedure' for CP Violation in 
\vspace*{2mm}
Dalitz Studies of $B$ (\& $D$ \& $\tau$) Decays}}

\vspace*{19mm}

{\bf I. Bediaga$^a$, I.I.~Bigi$^b$, A. Gomes$^c$, J. Miranda$^a$, J. Otalora$^c$, A. Reis$^a$ and A. Veiga$^d$} \\
\vspace{7mm}
$^a$ {\sl  Centro 
Brasileiro de Pesquisas F\'\i sicas, Rua Xavier Sigaud 150, 22290-180, Rio de Janeiro, RJ, Brazil}\\
$^b$ {\sl Department of Physics, University of Notre Dame du Lac}\\
{\sl Notre Dame, IN 46556, USA}\\
$^c$ {\sl Instituto de F\'{\i}sica, Universidade Federal do Rio de Janeiro}\\
$^d$ {\sl Departamento de Energia El\'etrica, Pontif\'{\i}cia Universidade Cat\'olica do Rio de Janeiro}
\vspace*{-.8mm}\\

{\sl email addresses: bediaga@cbpf.br, ibigi@nd.edu, jotalo@if.ufrj.br,
jussara@cbpf.br, alvaro@cbpf.br, alberto@cbpf.br} \\

\vspace*{10mm}

{\bf Abstract}\vspace*{-1.5mm}\\
\end{center}
The `Miranda Procedure' proposed for analyzing Dalitz plots for CP asymmetries in charged $B$ and 
$D$ decays in a model-independent manner is extended and refined. The complexity of CKM CP phenomenology 
through order $\lambda^6$ is needed in searches for New Dynamics (ND). 
Detailed analyses of three-body final states offer great advantages:  
(i) They give us more powerful tools for deciding whether an observed CP asymmetry represents 
the manifestation of ND and its features.   
(ii) Many advantages can already be obtained by the `Miranda Procedure' without construction of 
a detailed Dalitz plot description. (iii) One studies CP asymmetries 
independent of production asymmetries. 
We illustrate the power of a second generation  
Miranda Procedure with examples with time integrated rates for $B_d/\bar B_d$ decays to final states 
$K_S\pi^+\pi^-$ as trial runs 
with comments on $B^{\pm} \to K^{\pm}\pi^+\pi^-/K^{\pm}K^+K^-$.

\vspace{3mm}

\hrule

\tableofcontents
\vspace{5mm}

\hrule\vspace{5mm}

%%%%%%%%%%%%%
\section{Landscape of $B_{u,d,s}$, $D_{u,d,s}$ \& $\tau$ CP Violations}
%%%%%%%%%%%%

The predictions of CKM theory have been impressively confirmed to a degree that has 
persuaded a part of our community to focus on scenarios of Minimal Flavour 
Violation (MFV) -- i.e.,  models of New Dynamics (ND) that contain the same sources of flavour 
violations as the Standard Model (SM). Some intriguing work has been done along these lines, yet we view the 
hypothesis of MFV as far from compelling at present. The data still allow 
for sizable deviations from SM predictions in heavy flavour transitions; in $D^0$ and some $B_s$ decays 
they could even be dominant. Furthermore, baryogenesis requires the intervention of ND 
with CP violation. If ND appears around the $\mathcal{O}$(1 TeV) scale underlying the weak-electric phase transition,  
which is not intrinsically connected with 
flavour dynamics, it will affect CP asymmetries in heavy flavour decays -- but not on the leading level. 
CP asymmetries are given by three classes of observables, namely 
\begin{itemize}
\item 
$|q/p| \neq 1$, which shows purely indirect CP violation.  
\item 
Absolute amplitudes $|A_f| \neq |\bar A_f|$ that show purely direct CP violation and depend on the 
final states.   
\item 
The relative phases between $q/p$ and $\bar A_f \otimes A^*_f$, which will depend on the final state $f$. 
We write $\bar A_f \otimes A^*_f$ rather than just $\bar A_f  A^*_f$, because for a 
three-body final state one has to denote the position in the two-dimension plot. The significance of this feature  
will become clearer through the illustrations given below. 
\end{itemize}

%%%%%%%%%%%%%
\subsection{Present Status of CP Asymmetries in $B$, $D$ \& $\tau$ Transitions}
\label{DATA}
%%%%%%%%%%

Oscillations have been observed for all three heavy flavours $B_s$, $B_d$ and $D^0$ mesons, but on very different 
numerical levels. 

%%%%%%%%%%%%%
\subsubsection{Indirect CP Violation in $B_{d,s}$ Transitions}
\label{INDATABDS} 
%%%%%%%%%

Indirect CP violation has been measured with very good accuracy in $B_d \to  \psi K_S/K_L$ 
and $B_d \to \pi^+\pi^-$ \cite{pdg}: 
\bea
S(B_d \to \psi K_S) &=& 0.658 \pm 0.024
\\
S(B_d \to \pi^+\pi^-) &=& -0.61 \pm 0.08
\eea
Purely indirect CP violation gives $S(B_d \to \psi K_S) = - S(B_d \to \pi^+\pi^-)$. 

Very recent data from LHCb \cite{lhcb1} on $B_s \to \psi \phi$, $\psi f_0(980)$ give
\beq
\phi _s = - 0.002 \pm 0.083 \pm 0.027 \; {\rm rad}   
\eeq

%%%%%%%%%%%%%
\subsubsection{Direct CP Violation in $B_{u,d,s}$ Transitions}
\label{DIRDATABDS} 
%%%%%%%%%

Direct CP asymmetries has been established in $B_d$ decays  \cite{pdg, lhcb2}: 
\bea
A_{CP}(B_d \to K^+\pi^-)|_{\rm PDG'10} &=& - 0.098 \pm 0.013 \\ 
A_{CP}(B_d \to K^+\pi^-)|_{\rm LHCb '11} &=& - 0.088 \pm 0.011 \pm 0.008 \\
C(B_d \to \pi^+\pi^-) &=& - 0.38 \pm 0.17
\eea
No sign for direct CP violation has been found in $B_d \to K_S\pi^0$ -- by sizable asymmetry 
can still be allowed: 
\beq 
A_{CP}(B_d \to K_S\pi^0)|_{\rm PDG'10} = 0.00 \pm 0.13
\eeq
Intriguing evidences for direct CP violation has been found in $B_s$ and $B^+$ in 
(quasi-)two-body final states \cite{pdg,lhcb2,cdf1}: 
\bea
A_{CP}(\bar B_s \to K^+\pi^-)|_{\rm LHCb'11} &=&  + 0.27 \pm 0.08 \pm 0.02 \\
A_{CP} (B^+ \to D_{CP[+1]} K^+ ) &=& + 0.24 \pm 0.06 \\
A_{CP} (B^+ \to \rho^0 K^+ ) &=& + 0.37 \pm 0.10 \\ 
A_{CP} (B^+ \to f_0(1370) \pi^+ ) &=& + 0.72 \pm 0.22 \\
A_{CP} (B^+ \to \eta K^+ ) &=& - 0.37 \pm 0.09 \\
A_{CP} (B^+ \to f_2(1270) K^+ ) &=& - 0.68 ^{+0.19}_{-0.17} \\
A_{CP}(\bar B_s \to K^+\pi^-)|_{\rm LHCb'11} &=&  + 0.27 \pm 0.08 \pm 0.02 \\
A_{CP}(\bar B_s \to K^+\pi^-)|_{\rm CDF} &=&  + 0.39 \pm 0.15 \pm 0.08  
\eea

%%%%%%%%%%%%%
\subsubsection{Evidence for CP Asymmetries in $D^0$ Decays}
\label{INDATABDS} 
%%%%%%%%%

No sign of indirect CP violation has been found in $D^0 \to K^+K^-/\pi^+\pi^-$ \cite{HFAG}: 
\beq
\left |\frac{q}{p}  \right | = 0.88 ^{+0.18}_{-0.16} \; , \;  
\phi   = \left( - 10.2 ^{+9.4}_{-8.9}\right) ^o
\eeq

Direct CP asymmetry has been found in $\Gamma (D^0 \to K^+K^-) - \Gamma (D^0 \to \pi^+\pi^-)$ with 
3.5 sigma away from zero by LHCb and with 2.7 sigma by CDF;   
\bea
\Delta A_{CP} &=&   
- 0.82 \pm 0.21({\rm stat}) \pm 0.11({\rm syst}) \% \;\; {\rm LHCb} \cite{dacp} \\
\Delta A_{CP} &=&  
- 0.62 \pm 0.21({\rm stat}) \pm 0.10({\rm syst}) \% \;\; {\rm CDF} \cite{CDF4}
\eea
with $\Delta A_{CP}\equiv A_{CP} (D^0 \to K^+K^-) - A_{CP} (D^0 \to \pi^+\pi^-)$. 
This is the first significant evidence for $CP$ violation 
in $\Delta C \neq 0$ dynamics, and it is important whether it is due to alone SM or need 
impact from ND.

%%%%%%%%%%%%%
\subsubsection{Evidence for CP Asymmetries in $\tau$ Decays}
\label{INDATAU} 
%%%%%%%%%

In $\tau ^- \to K_S\pi^- \nu$ decays one has a prediction \cite{BSLEP}
\beq
A_{\rm CP} (\tau ^+ \to \bar \nu + K_S\pi^+ )|_{\rm SM} = ( 0.36 \pm 0.01)\%  \; , 
\eeq
independent of dynamics that generate $K^0 \to \bar K^0$ oscillations, 
and data from the BaBar Collab. \cite{BABARTAU}: 
\beq
A_{\rm CP} (\tau ^+ \to \bar \nu + K_S\pi^+[\geq 0 \, \pi^0 ] )|_{\rm BaBar} = (- 0.36 \pm 0.23 \pm 0.11)\% \; .
\eeq

%%%%%%%%%%%%%
\subsection{CKM Matrix Parametrization through ${\cal O}( \lambda ^6)$}
\label{CKM6}
%%%%%%%%%%%%

The usually applied the Wolfenstein parameterization of the CKM matrix gives real parts through 
${\cal O}(\lambda^3)$ and the imaginary part through ${\cal O}(\lambda^4)$. The CKM matrix is usually 
described by the four parameters $\lambda$, $\rho$, $\eta$ and $A$ with the last three 
ones of order unity;  thus one gets 
$|V_{ub}/V_{cb}| \simeq \lambda \sqrt{\rho^2 + \eta^2} \sim {\cal O}(\lambda)$.
 
PDG states $|V_{ub}/V_{cb}| \sim 0.085$. The global fit leads to $\rho \simeq 0.13$ and $\eta \simeq 0.34$. 
It means that one has to use a parametrization with through order of $\lambda^6$ and with other quantities 
of true order of unity. One has been found in Ref.\cite{AHN}:
\begin{eqnarray} 
\left(
\begin{array}{ccc}
 1 - \frac{\lambda ^2}{2} - \frac{\lambda ^4}{8} - \frac{\lambda ^6}{16}, & \lambda , & 
 \bar h\lambda ^4 e^{-i\delta_{\rm QM}} , \\
 - \lambda + \frac{\lambda ^5}{2} f^2,  & 
 1 - \frac{\lambda ^2}{2}- \frac{\lambda ^4}{8}(1+ 4f^2) 
 -f \bar h \lambda^5e^{-i\delta_{\rm QM}}  &
   f \lambda ^2 +  \bar h\lambda ^3 e^{-i\delta_{\rm QM}}   \\
    & +\frac{\lambda^6}{16}(4f^2 - 4 \bar h^2 -1  ) ,& -  \frac{\lambda ^5}{2} \bar h e^{-i\delta_{\rm QM}}, \\
 f \lambda ^3 ,&  
 -f \lambda ^2 -  \bar h\lambda ^3 e^{-i\delta_{\rm QM}}  & 
 1 - \frac{\lambda ^4}{2} f^2 -f \bar h\lambda ^5 e^{-i\delta_{\rm QM}}  \\
 & +  \frac{\lambda ^4}{2} f + \frac{\lambda ^6}{8} f  ,
  &  -  \frac{\lambda ^6}{2}\bar h^2  \\
\end{array}
\right)
+ {\cal O}(\lambda ^7)
\end{eqnarray}
A {\em global} fit of the CKM matrix gives: $\lambda \simeq 0.225$, $f \simeq 0.75$, $\bar h \simeq 1.35$ 
and the `maximal' phase $\delta_{\rm QM} \simeq 90^o$. 

This pattern is not so obvious as from the Wolfenstein parametrization, more subtle for 
CP violation and is similar only in a semi-quantitive way \cite{BIPAUL}.  To give three examples: 
\begin{itemize}
\item
CP asymmetry in   
$B_d \to \psi K_S$ depends in SM on 
\beq
-{\rm Im}\frac{V^*_{tb}V_{td}}{V_{tb}V^*_{td}} \frac{V_{cb}V^*_{cs}}{V^*_{cb}V_{cs}}    
\simeq \frac{\frac{2\bar h\lambda }{f }{\rm sin}\delta_{\rm QM} +
\left( \frac{\bar h \lambda }{f } \right)^2 {\rm sin}2\delta_{\rm QM}   }{ 1+ \left( \frac{\bar h \lambda }{f } \right)^2 + 
\frac{2\bar h\lambda }{f }{\rm cos}\delta_{\rm QM}  }
\eeq
One gets: 
\bea
S(B_d \to \psi K_S)= {\rm sin}2\phi_1 &\simeq& 0.63 - 0.69   \; \; {\rm for} \; \delta _{\rm QM} \simeq 75^o - 90 ^o \\
S(B_d \to \psi K_S)= {\rm sin}2\phi_1 &\sim& 0.74   \; \; {\rm for} \; \delta _{\rm QM} \simeq 100^o - 120^o \; ; 
\eea
i.e., CKM dynamics could produce 
$S(B_d \to \psi K_S) \sim 0.74$ as largest value for CP asymmetry with $\delta _{\rm QM} \simeq 100^o - 120^o$, 
not with the maximal $\delta _{\rm QM} = 90 ^o$. 

\item
Again one finds that indirect CP violation in $B_s$ is CKM suppressed in the SM by 
\beq
{\rm Im}\left[ \frac{V^*_{tb}V_{ts}}{V_{tb}V^*_{ts}}\frac{V_{cb}V^*_{cs}}{V^*_{cb}V_{cs}}\right] \simeq   
\frac{2(\bar h /f)\lambda^3\, [ {\rm sin}\delta_{\rm QM} +2(\bar h /f){\rm sin}2\delta_{\rm QM}  ]}
{1+(4\bar h/f)\lambda \, {\rm cos}\delta_{\rm QM}} 
\sim 0.03 - 0.05 \; .  
\eeq
with $\delta _{\rm QM} \simeq (75 - 120) ^o$. 

\item
Direct CP violation in $B^{\pm} \to D_+ K^{\pm}$ depends on sin$\phi_3$, where  one gets:  
\beq
\phi_3 = {\rm arg}\left( \frac{V^*_{ub}V_{ud}}{-V^*_{cb}V_{cd}}\right) \simeq 
(1-\lambda^2/2) \frac{\bar h\lambda}{f} \frac{{\rm sin}\delta _{\rm QM} }
{1+(\bar h \lambda/f )^2 + 2(\bar h\lambda/f) {\rm cos}\delta _{\rm QM}}
\eeq
Thus  
\beq
\phi_3 = 0.28 \; / \; 0.34 \; / \; 0.42  \; \; \; {\rm for} \; \; \; \delta_{\rm QM} = 75^o \; / \; 90^o \; / \; 110^o \; .  
\eeq

\end{itemize}
Therefore sin$2\phi_1\simeq 0.69 \pm 0.06$ and sin$\phi_3 \simeq 0.34 \pm 0.07$ are consistent with CKM 
dynamics with {\em lower} values of $\phi_1$ \& $\phi_3$ correlated with each other with  ~10 \% vs. ~20 \%. 
Nevertheless the impact of ND can `hide' in predicted CP asymmetries.

%%%%%%%%%%%%
\subsection{Present Resume on CP Asymmetries in Two-Body Final States in $B$ and $D$ Decays}
%%%%%%%%%%%

SM generates indirect and direct CP asymmetries in $B$ \& $D$ (and in $K$) transitions. 
Their strengths are based on several items: 
\begin{itemize}
\item 
The CKM matrix is discussed above in Sect.\ref{CKM6}. 
We can say that SM is at least the leading source of 
CP violation in $B$ on most transitions -- except at present for $B_s \to \psi \phi/\psi f_0(980)$. 
On the other hand ND can affect CP asymmetries on the level of $\sim 10 - 20 \%$ for $B_{u,d}$ decays.   
Furthermore one has to focus on correlations with CKM suppressed decays of  $B_{u,d,s}$ (and $K$ and 
$D_{(s)}$) on the level of 20 \%.  Therefore one need more accuracy from data and their 
interpretation to find impact of one (or two) ND -- and to probe three-body final states. 

\item
While the final states $K^+\pi^-$ in $B_d$ and $\bar B_s$ are the same, the underlying dynamics  
are very different: 
\begin{itemize}
\item 
SM amplitudes for $\bar B_d \to K^-\pi^+$ are given by `tree' Cabibbo 
suppressed transitions $b \to u \bar u s$ and (1-loop) `Penguin' $b \to s \bar q q$; 
$\sim$ 10 \% CP asymmetry seems a reasonable value in SM. 

\item 
On the other hand SM amplitude for $\bar B_s \to K^+\pi^-$ is given by `tree' 
Cabibbo favoured $b \to u \bar u d$ and Cabibbo suppressed `Penguin' 
$b \to d \bar q q$. Therefore one expects CP asymmetry on the `natural' level of 
${\cal O}(1\% )$. 

ND can enhance `Penguin' amplitude significantly. However one expects such effect in other 
$B$ decays -- unless a nearby resonance can affect mostly $B_s$, but not $B_{d,u}$ decays. 

\end{itemize}

For $D$ decays the landscape is much more subtle, but also very `topical': 
\begin{itemize}
\item 
CKM dynamics produce direct CP asymmetries 
in singly Cabibbo suppressed (SCS) decays around the scale of 0.001. 
\item 
CP asymmetries in double 
Cabibbo suppressed (DCS) ones are zero at $\mathcal{O}(\lambda ^4)$.   
\end{itemize}

\item
CP asymmetries are controlled by non-perturbative QCD.  

\item 
The difference between $A_{\rm CP} (\tau ^+ \to \bar \nu + K_S\pi^+[\geq 0 \, \pi^0 ] )|_{\rm measured}$ and  
$A_{\rm CP} (\tau ^+ \to \bar \nu + K_S\pi^+ )|_{\rm SM}$ depend on 
our control of non-perturbative QCD -- like the plots of the final states of 
$A_{\rm CP} (\tau ^+ \to \bar \nu + [K\pi ]^+ )$ and $A_{\rm CP} (\tau ^+ \to \bar \nu + [K\pi\pi ]^+ )$
\cite{TAU}.

\end{itemize}  

Finding impact of ND in CP asymmetries in $B$ and $D$ decays is one thing -- 
however probing the `shape' of one (or two) ND is another challenge.   

Dedicated studies of three-bodies final states are needed to identify 
important features of ND involved \cite{MIRANDA1}. 
Three-body final states analyses are very time consuming. 
In Sect.\ref{ADV} we list the general advantages that such analyses merit the needed work; we first sketch 
in Sect.\ref{STAGE} the situations for three-body decays of $B_s$, $B_d$, $D^0$ in general for searching 
CP asymmetries through (partly) time integrated data to set the stage; afterwords we give more realistic situations 
in $B_d$ transitions in 
Sect.\ref{REALBD} and comments on $B_s$ transitions in Sect.\ref{REALBS}; 
finally we summarize our main conclusions in Sect.\ref{CON}.

%%%%%%%%%%%%
\section{Advantages of Studies of Three-Body Final States}
\label{ADV}
%%%%%%%%%%%

While the weak dynamics from CKM and ND are the driving forces for CP asymmetries, one has to 
control FSI from non-perturbative QCD not only in a qualitatively way -- that is the focus of our study. 

%%%%%%%%%%%%%
\subsection{Opportunities Offered by Dalitz Plot Studies}
\label{OPPORT}
%%%%%%%%%%%%

No study of any three-body final states in $B$ decays have found an established CP violation, and none from 
$K$ or $D$ mesons shows any sign for it. However there  
can be -- actually they are more likely to be found. {\em The average over CP asymmetries in a Dalitz plot is 
expected to be much larger than `local' asymmetries, which often compensate with each other.} 
As explained in some detail in \cite{MIRANDA1}, 
crucial insights into CP odd dynamics 
will be learnt from their impact on final state distributions. Dalitz studies will play 
a central role in the future for several reasons: 
\begin{itemize} 
\item 
Differential or `local' asymmetries could be 
considerably larger than ones averaged over the Dalitz plot. 
\item 
For two-body final states there is only one CP asymmetry, namely $\Gamma (B^0/D^0 \to h^+h^-)$ vs. 
$\Gamma (\bar B^0/\bar D^0 \to h^+h^-)$.  On the other hand the topologies of Dalitz plots for $B^0 \to h^+ h^-h^0$ 
and $\bar B^0 \to h^+h^-h^0$ are in general different; for example the two half of the plots 
$s_{h^+h^0} - s_{h^-h^0}$ for the $B^0$ and $\bar B^0$ are different. {\em However their sum have to 
be symmetric -- unless CP asymmetries occur!}  

\item 
While the difference in total rates for $B/D \to 3h$ vs. $\bar B/\bar D \to 3h$ are affected by production asymmetries, 
differences between corresponding regions in the Dalitz plots are not. 

\item 
Nontrivial correlations provide powerful 
validation tools. 
\item  
The pattern of a CP asymmetry that has emerged in a Dalitz plot 
can tell us about the spin structure of the underlying effective operator. 
\item 
The cleanest experimental sign whether an observed asymmetry is produced by direct or indirect CP violation 
(or which parts are due to one or the other) is their dependence on the time of decay. Direct 
asymmetry is independent of the time of decay, whereas indirect violation evolutes in time in a clear prescription, 
since it is driven by oscillations. With only time integrated data with two-body final states one cannot decide it. 
If one observes a 
CP asymmetry in a leading CKM final state -- like $B_d \to \psi K_S$ or $B_s \to \psi \phi$ -- you will 
argue that it is most likely indirect. Yet for CKM-suppressed decays you hardly have such an argument. 
As explained later, one can use more decision criterions from three-body final states.

\item 
Time {\em depending} CP asymmetries give us more informations about underlying dynamics for 
$B^0$ and $D^0$ transitions. Of course one needs more statistics. {\em Partially time integrated} rates for 
three-body final states give us more insights. 
\end{itemize}

Dalitz studies offer also a more technical advantage when searching for CP asymmetries, 
namely `tunable' strong phases. Since that is of direct relevance for our subsequent analysis, 
we explain it next. 

%%%%%%%%%%%%%%%
\subsection{Phases with Breit-Wigner Resonances}
\label{BWPHASES}
%%%%%%%%%%%%%%

With CP violation being expressed by a complex weak phase due to CPT invariance, it can lead to an 
observable asymmetry only if one has the interference between two different amplitudes. 
Yet more is needed: hadronization has to affect the two amplitudes differently. This is usually 
expressed by stating that the two amplitudes have to exhibit 
{\em different weak as well as strong} phases. 

Consider the decay $P \to f$ receiving contributions from two coherent amplitudes,
\bea
A(P\to f) &=& e^{i\phi_1^W}\ e^{i\delta_1^{\mathrm{FSI}}}|\mathcal{A}_1| +
e^{i\phi_2^W}\ e^{i\delta_2^{\mathrm{FSI}}}|\mathcal{A}_2| \\
A(\bar P\to \bar f) &=& e^{-i\phi_1^W}\ e^{i\delta_1^{\mathrm{FSI}}}|\mathcal{A}_1| +
e^{-i\phi_2^W}\ e^{i\delta_2^{\mathrm{FSI}}}|\mathcal{A}_2|, 
\eea
where $\phi_i^W$ and $\delta_i^{\mathrm{FSI}}$ are the weak and strong phases, respectively, and
$\mathcal{A}_i$ are the moduli of the amplitudes. The CP asymmetry between partial widths 
\beq
A_{\mathrm{CP}} = \frac{\Gamma (P \to f) - \Gamma (\bar P \to \bar f)}
{\Gamma (P \to f) + \Gamma (\bar P \to \bar f)} , 
\eeq
is given by 
\beq
A_{\mathrm{CP}} = \frac{2\sin(\Delta\phi^W) \sin(\Delta\delta^{\mathrm{FSI}}) |\mathcal{A}_2\mathcal{A}_1|}
{1 + |\mathcal{A}_2\mathcal{A}_1|^2 + 2|\mathcal{A}_2\mathcal{A}_1|\cos(\Delta\phi^W) 
\cos(\Delta\delta^{\mathrm{FSI}})}
\label{acp}
\eeq 
CP violation is induced by $\Delta\phi^W$, but it becomes observable only if the final state
interaction (FSI) introduces a non-trivial phase shift.

For two-body final states it is often implied that the strong 
phases $\delta^{\mathrm{FSI}}_i$ carry a {\em fixed} value for a given final state $f$ (the two 
amplitudes are assumed to have different isospin contents up to isospin violation).

In the case of three-body decays, the transition $P\to f$ is dominated by resonant intermediate states.    
The requirement of non-trivial strong phase different is satisfied by the energy dependent phases
of the resonances. The Breit-Wigner excitation curve for a resonance $R$ reads 
\beq 
F^{\rm BW}_{R} (s) = \frac{1}{m^2_{R} - s - i m_{R}\Gamma_{R}(s)}  ~,
\label{BW}
\eeq 
introducing a sizable phase as expressed through 
\beq 
{\rm Im}F^{\rm BW}_{R} (s) = \frac{m_{R}\Gamma_{R}(s)}{(m^2_{R} - s)^2+
(m_{R}\Gamma_{R}(s))^2 } \; , 
\label{IMBW}
\eeq 
where $\Gamma_{R}(s)$ denotes the energy dependent relativistic width. 

For $P\to p_1p_2p_3$ we define $s_1 = (p_1+p_2)^2$ and $s_2 = (p_1+p_3)^2$.
The previously constant strong phases and amplitude moduli in Eq.(\ref{acp}) now depend on the position in the
Dalitz plot, $\delta^{\mathrm{FSI}} \to \delta(s_1,s_2)$ and
$\mathcal{A}\to \mathcal{A}(s_1,s_2)$. The resonant amplitudes populate the whole
phase space in $D$ decays, and a large portion of it in $B$ decays. Therefore, the CP asymmetry 
will also depend on the Dalitz plot coordinates, $A_{\mathrm{CP}} \to A_{\mathrm{CP}}(s_1,s_2)$.

After taking the modulus square of these amplitudes one reads off that a CP asymmetry will 
arise, when there are non-zero {\em weak} phases. Having to deal with non-uniform strong phases might 
appear as a complication that just creates more work for analysis of decays with three-(and four-)body 
final states. However there is an award for extra works: a resonance -- in particular if it is 
relatively narrow -- can tell us where 
CP asymmetries have to be and that the asymmetry has to change over a relatively narrow range in 
the Dalitz plot. 
The Dalitz plots carry the same area independent of production asymmetries; yet the 
{\em relative corresponding population 
densities} probe CP invariance.

%%%%%%%%%%%%%%%
\subsection{`Miranda Procedure'}
\label{MIRPRO}
%%%%%%%%%%%%

{\em Indirect} CP violation in $B^0$ decays is and will be well measured in 
$B_d \to \psi K_S$ and $B_s \to \psi \phi , \psi f_0(980)$ -- and maybe in $D^0 \to \phi K_S$;  
ND can impact those transitions in a sizable way. {\em Direct} CP violations affect final states in different 
strengths, in particular CKM suppressed ones both in SM and ND. The {\em existence} of ND might -- and probably 
will -- be found in indirect and direct CP violation in two-body final states -- 
yet its main {\em features} have to extracted from three-(and four-)body final states. Some asymmetries 
can tell us about the spin operators creating about etc. There ere three classes of sources of 
CP asymmetries, namely 
\begin{enumerate}
\item
from CP conjugated quasi-two-body final states;  
\item
interference between quasi-two-body final states; 
\item
contributions from "true" three-body final states or broad resonances like $\sigma$. 
\end{enumerate}
Contributions from the first class like $K\rho$ or $\pi \rho$ are obvious. 
However from the second class one finds positive and negative contributions to the CP asymmetry; 
therefore those get washed out from the total integral over the phase space. This applies mostly to 
the third class of CP asymmetries. Therefore we will denote CP asymmetries from the first and second classes 
by $CPV_A$ and $CPV_I$ below. For the third class one can help sizably from future theoretical efforts. 

The advantages listed above for CP studies in three-body final states justify 
the considerable `overhead' in statistics and tuning efforts that constructing a 
satisfactory Dalitz model requires. Yet one can{\em not count}  
on obtaining a {\em unique} Dalitz model even with infinite statistics. It is thus of considerable practical value to 
develop another method for analyzing a Dalitz plot that is model-independent; it will allow important statements about 
the existence  of a CP asymmetry and its approximate localization inside the Dalitz plot with smaller data sets than 
constructing a full fledged Dalitz model. At the very least it would help to identify the sub-domain in the Dalitz 
plot where one had to focus the tuning efforts for the Dalitz model. 

In Ref. \cite{MIRANDA1} we have proposed one method that can serve such a purpose in a quantifying 
way. When searching for CP asymmetries we have suggested to analyze the {\em significance} 
\beq
\Sigma (i) \equiv \frac{N(i)- \bar N(i)}{\sqrt{N(i) + \bar N(i)} } \; , 
\label{SIGMA}
\eeq
which amounts to a standard deviation for a Poissonian distribution rather 
than the customary {\em fractional} asymmetry 
\beq
\Delta (i) \equiv \frac{N(i) - \bar N(i)}{N(i) + \bar N(i)}
\label{DELTA}
\eeq 
in particle vs. anti-particle populations $N(i)$ and $\bar N(i)$ 
for each bin $i$, respectively.  One analyzes, whether the $\Sigma (i)$ distribution has a higher 
frequency of exhibiting large deviations from zero than expected for fluctuations 
We have illustrated this method -- `mirandizing' in vernacular -- 
by applying it to $B^{\pm} \to K^{\pm} \pi^+\pi^-$ and $D^{\pm} \to \pi^{\pm}\pi^+\pi^-$;  the Dalitz plots have 
been constructed with fast MC simulations making specific assumptions about the 
underlying dynamics and the source of the CP asymmetry. In those pilot studies we could show 
that using the observable $\Sigma (i)$ instead of $\Delta (i)$ allowed a much more robust extraction of the seeded CP 
asymmetry and its location inside the Dalitz plot. It remains to be seen of course 
how mirandizing holds up when treating real data.  

Our claim is {\em not} to replace full fledged Dalitz plot analysis. Our goal is to present an analysis that 
can produce significant results on CP violations from smaller data set while maintaining many of the 
advantages of a full Dalitz plot study. The latter is still the final goal of our road to `Rome'  -- the 
impact of New Physics on CP violations in nonleptonic decays of beauty and charm hadrons. We also want to encourage 
others to try other possible roads to this `Rome'. An interesting work can be found in Ref. \cite{Williams}

%%%%%%%%%%%%%%%
\subsection{Comment on CP Asymmetries in $\tau^- \to \nu [K\pi/K\pi\pi ]^-$}
\label{TAUMIRANDA}
%%%%%%%%%%%

The SM generates a global CP asymmetries in $\tau ^- \to \nu K_S [S=0]$ due to $K^0 - \bar K^0$ oscillations 
with 2Re $\epsilon_K$, but not beyond that. On the other hand ND has a larger chance to appear in the 
CP asymmetries in $\tau ^- \to \nu [K \pi 's]^-$, since SM amplitudes are Cabibbo suppressed. 

For $\tau ^- \to \nu [K \pi]^-$ one has a three-body final states in general, and in 
$\tau ^- \to \nu [K2\pi/3K]^-$ one has three-body hadronic final states. The `Miranda Procedure' can and should 
be applied here with some refining Dalitz plots: the masses of the hadronic systems are not fixed as in $B$ and $D$ 
decays.

%%%%%%%%%%
\section{Setting the Stage for Probing CP Invariance}
\label{STAGE}
%%%%%%%%% 

Both indirect and directly CP violations have been established in $K^0$ and $B_d$ transitions, and 
SM through CKM theory gives at least the leading contributions. The ND expected -- or hoped for -- to find around 
the few TeV scales should produce some `footprints' through CP asymmetries in $B$ and $D$ transitions. 
However `reading' them produces large both experimental and theoretical challenges. The time evolutions of the 
flavour tagged transitions provide the most powerful tool in identifying to source of the underlying 
dynamics. To begin this projects we want to show what you learn from {\em non-flavour tagging} transitions 
with{\em out} time resolved analyses. 

For this study we describe the time integrated rates also for non-flavour tagged 
data and give comments on partly time resolved rates. 

CP asymmetries are control by five observables:  
\begin{enumerate}
\item 
$x=\Delta M/\bar \Gamma$ and $y=\Delta \Gamma /2\bar \Gamma$, which are insensitive to CP violation;  
\item 
$|q/p| \neq 1$, which shows purely indirect CP violation, and is determined by 
$|q/p| \simeq 1 -\frac{1}{2} a^{\rm CP}_{SL}$, where denote the CP asymmetry in semi-leptonic decays in 
`wrong'-sign leptons;  
\item 
absolute amplitudes $|A_f| \neq |\bar A_f|$ that show purely direct CP violation and depend on the 
final states and 
\item 
the relative phases between $q/p$ and $\bar A_f \otimes A^*_f$, which will depend on the final state $f$ due to direct 
CP violation. We write $\bar A_f \otimes A^*_f$ rather than just $\bar A_f  A^*_f$, because for a 
three-body final state one has to denote the position in the two-dimension plot. The significance of this feature  
will become clearer through the our illustrations later. 
\end{enumerate}
We give general expressions with these observables. Then we show that in describing for $B_d$ decays we 
can ignore $y_d$ effects, while for $D^0$ decays one has to include both $x_D$ and $y_D$ dependences, 
but only to first order. For $B_s$ transitions one has $x_s \gg y_s$, but one has to include $y_s$ effects due to 
spectacularly fast $x_s$ oscillations for time integrated data; CP asymmetries controlled by 
Im$\frac{q_s}{p_s}\bar A_f \otimes A_f$ are suppressed by $1/x_s$. 

Indirect CP violation affects all channels through two quantities, namely $|q/p|$ and the relative phase 
between $q/p$ and $\bar A_f \otimes A^*_f$; their weight of course depends on $\left| |q/p| - 1\right|$ and 
the strength of $|\bar A_f \otimes A^*_f|$. As mentioned above we can use the approximation of $|q/p|=1$ for 
$B_d$ and $B_s$ channels; the effect of indirect CP violation is affected the direct CP on rate $B_d$ and 
$B_s$ modes and therefore the impact of ND that is probably different for mode to mode. 

For two-body final states like 
$B_s \to h^+h^-$ vs.  $\bar B_s \to h^+h^-$  time dependent CP asymmetries are reduced by $1/x_s^2$; 
however for direct CP violation in 
$B_s \to h^+h^-h^0$ vs. $\bar B_s \to h^+h^-h^0$ one can find an asymmetry between corresponding 
regions of the {\em sum of Dalitz plots of $B_s \to f$ and $\bar B_s \to f$ due to interference effects} -- i.e., 
with{\em out} flavour tagging.  

Obviously flavour tagged time resolved analyses bring the largest information about the underlying 
dynamics; our main goal for this study is how many lessons can be obtained from non-flavour tagged time 
integrated data. Flavour-tagged and time resolved data will come later.

%%%%%%%%%%%%%
\subsection{Three-Body Decays for Neutral Mesons}
\label{GENERALTHREEB}
%%%%%%%%%%%%

For neutral $B$ or $D$ decays into two-body final states like $K^+K^-$ or $\pi^+\pi^-$, it is clear that it 
is a (even) CP eigenstate. For $f = h^+h^-h^0$ like $K_S\pi^+\pi^-$, $K_SK^+K^-$ or 
$\pi^+\pi^-\pi^0$ the judgement is more complex: it can be [CP=+] 
$K_Sf_0(980)/K_S\sigma \to K_S\pi^+\pi^-$, [CP=-] $K_S\phi \to K_SK^+K^-$,  
[CP=-] $\rho^0\pi^0 \to \pi^+\pi^-\pi^0$ or [CP=+] $\sigma \pi^0 \to \pi^+\pi^-\pi^0$. At the same time one has 
final states $K^{*\pm}\pi^{\mp}$, $\rho^{\pm}\pi^{\mp}$ etc. and interferences between them. Not only the  
total time integrated widths for $P \to h^+h^-h^0$ vs. $\bar P \to h^+h^-h^0$ give us a lesson on CP violating 
dynamics, but also their `topologies' -- i.e., the distributions over the Dalitz plots. Therefore we denote 
$A_f$ and $\bar A_{\bar f}$ for $P\to h^+h^-h^0$ and $\bar P\to h^+h^-h^0$, respectively.

The `Miranda procedure' can be applied to all three-body final states, but here we will discuss it only for 
$f = h^+h^-h^0$ like $K_S\pi^+\pi^-$. 

The {\em time dependent} rates can distinguish the three types of CP violations: 
$|q/p| \neq |p/q|$, $|A_f| \neq |\bar A_{\bar f}|$ and 
Im$\frac{q}{p}\bar A_{\bar f} \otimes A_f^*$ $\neq$ Im$\frac{p}{q}A_f \otimes \bar A_{\bar f}^*$. 
For practical reasons we focus on {\em time integrated} widths for this study: 
\bea
\Gamma ( P \to f) = \frac{C}{\Gamma_1} \left[ a + \frac{1}{1 - \frac{\Delta \Gamma}{\Gamma_1}}b
+ \frac{1}{1 - \frac{\Delta \Gamma}{2\Gamma_1}}
\frac{1}{1+ \frac{(\Delta M)^2}{(\Gamma_1 - \frac{\Delta \Gamma}{2})^2}} 
\left[ c + \frac{\Delta M}{\Gamma_1 - \frac{\Delta \Gamma}{2}}d  \right] \right]
\\
\Gamma ( \bar P \to \bar f) = \frac{C}{\Gamma_1} \left[ \bar a + \frac{1}{1 - \frac{\Delta \Gamma}{\Gamma_1}}\bar b
+ \frac{1}{1 - \frac{\Delta \Gamma}{2\Gamma_1}}
\frac{1}{1+ \frac{(\Delta M)^2}{(\Gamma_1 - \frac{\Delta \Gamma}{2})^2}} 
\left[ \bar c + \frac{\Delta M}{\Gamma_1 - \frac{\Delta \Gamma}{2}}\bar d  \right] \right]
\eea
CP violation can appear in the {\em time integrated} widths in principle from the three sources listed above. 

While the time {\em integrated} observables have the largest statistics, the time {\em dependent} ones 
have most detailed information about the underlying dynamics; {\em partially} integrated one can give us 
most of that information depending on how the parameters for the hadrons and their transitions.   
\bea
\Gamma _t^{\infty} \equiv \int_t^{\infty} dt \Gamma (P \to f; t) 
\\
\bar \Gamma _t^{\infty} \equiv \int_t^{\infty} dt \Gamma (\bar P \to \bar f; t) \; ; 
\eea
therefore 
\bea
\Gamma_0^t = \Gamma (P \to f) -  \Gamma _t^{\infty} 
\\
\bar \Gamma_0^t = \Gamma (\bar P \to {\bar f}) -  \bar \Gamma _t^{\infty}
\eea

%%%%%%%%%%%%%%%%%%%%%
\subsubsection{$B_d \to h^+h^-h^0$}
\label{GENBD}
%%%%%%%%%%

For $B_d$ transitions one can simplify these expressions in two ways, namely 
$y_d \simeq 0$ and $|q_d/p_d| \simeq 1$ for realistic experimental sensitivities.  
Integrated over {\em all times of decays} we get
\bea
\int _0^{\infty}dt|{\cal A}(B_d   \to f;t)|^2  = \frac{C}{\Gamma_{B_d}}  
\left[  |A_f|^2 + |\bar A_{\bar f}|^2 +  \frac{1}{1+x_d^2} \left( |A_f|^2 - |\bar A_{\bar f}|^2 \right) 
 -  \frac{2x_d}{1+x_d^2} {\rm Im}\left( \frac{q_d}{p_d}  \bar A_{\bar f} \otimes A^*_f\right) \right]
 \\ 
 \int _0^{\infty}dt|{\cal A}(\bar B_d   \to {\bar f};t)|^2  = \frac{C}{\Gamma_{B_d}}  
\left[ |A_f|^2 + |\bar A_{\bar f}|^2 +  \frac{1}{1+x_d^2} \left( |\bar A_{\bar f}|^2 - |A_f|^2 \right) 
 +  \frac{2x_d}{1+x_d^2} {\rm Im}\left( \frac{q_d}{p_d}  \bar A_{\bar f} \otimes A^*_f\right)  \right]
\eea
For equal productions of $B_d$ and $\bar B_d$ -- $\rho (B_d) = \frac{1}{2} = \rho (\bar B_d)$ -- we get 
\beq
\frac{1}{2} \left[ \int _0^{\infty}dt|{\cal A}(B_d   \to f;t)|^2 
+ \int _0^{\infty}dt|{\cal A}(\bar B_d   \to f;t)|^2\right] = |A_f|^2 + |\bar A_{\bar f}|^2
\label{BDNOPROD}
\eeq
For a two-body final state $f = h^+h^-$ one gets no information about CP violation from the 
time integrated sum as expected. However for $f = h^+h^-h^0$ direct CP violation can produce an 
asymmetry in {\em corresponding} regions of the {\em sum of Dalitz plots} as sketched before. 
Our studies given below will illustrate this feature. Their sum is weighted by their ratio due to a real production 
asymmetry (or a difference in their 
efficiencies) 
\bea
\rho (B_d)\int _0^{\infty}dt|{\cal A}(B_d   \to f;t)|^2 
+ (1-\rho (B_d))\int _0^{\infty}dt|{\cal A}(\bar B_d   \to {\bar f};t)|^2 = 
\nonumber \\
\frac{1}{1+x_d^2} \left[  (2\rho(B_d) + x_d^2)|A_f|^2 +(2(1-\rho(B_d)) +x_d^2)|\bar A_{\bar f}|^2 
+ 2x_d (1-2\rho (B_d)) {\rm Im}\left( \frac{q_d}{p_d}  \bar A_{\bar f} \otimes A^*_f\right)\right] 
\label{BDPROD}
\eea

If there is {\em production asymmetry}, it is not a `vice', but a `virtue'. It can be tracked by 
$\bar B_d \to \psi K^-\pi^+$ vs. $B_d \to \psi K^+\pi^-$ or by $B^{\pm} \to \psi K^{\pm}$. The 
strength of indirect CP violation is measured in $B_d \to \psi K_S$, whether the SM produces 
the whole or just the leading source of it. Used as an {\em input}  
for $B_d \to K_S\pi^+\pi^-$, $K_SK^+K^-$ one can interpret the impact of direct CP violation 
through the term Im$\left( \frac{q_d}{p_d}  \bar A_{\bar f} \otimes A^*_f\right)$, see Eq.(\ref{BDPROD}).

%%%%%%%%%%%%%%%%%%%%%
\subsubsection{$B_s \to h^+h^-h^0$}
\label{GENBS}
%%%%%%%%%%

For $B_s$ transitions, one can assume $|q_s/p_s| \simeq 1$, since even with sizable ND contributions to 
$B_s - \bar B_s$ $|q_s/p_s$ can differ from unity not more than several permil. While $\Delta \Gamma_s$ is small -- 
$y_s = \Delta \Gamma_s/2\bar \Gamma_s \simeq 0.094 \pm 0.024$ -- it should not been ignored.  
Integrated over all times of decays we get
\bea
\int _0^{\infty}dt|{\cal A}(B_s   \to f;t)|^2  \propto \frac{1}{2\Gamma_1} \cdot &&
\nonumber \\
\left[ |A_f|^2 + |\bar A_{\bar f}|^2 + \frac{\Delta \Gamma_s}{\Gamma_1}
\left( \frac{1}{2} ( |A_f|^2 + |\bar A_{\bar f}|^2)  - {\rm Re}\left( \frac{q_s}{p_s}\bar A_{\bar f} \otimes A^*_f  \right)  \right)
- \frac{2}{x_s}{\rm Im}\left( \frac{q}{p} \bar A_{\bar f} \otimes A^*_f\right) + {\cal O}(1/x_s^2)
\right]
 \\
\int _0^{\infty}dt|{\cal A}(\bar B_s   \to {\bar f};t)|^2  \propto \frac{1}{2\Gamma_1} \cdot &&
\nonumber \\
\left[ |A_f|^2 + |\bar A_{\bar f}|^2 + \frac{\Delta \Gamma_s}{\Gamma_1}
\left( \frac{1}{2} ( |A_f|^2 + |\bar A_{\bar f}|^2)  - {\rm Re}\left( \frac{q_s}{p_s}\bar A_{\bar f} \otimes A^*_f  \right) \right)
+ \frac{2}{x_s}{\rm Im}\left( \frac{q}{p} \bar A_{\bar f} \otimes A^*_f\right) + {\cal O}(1/x_s^2)
\right]
\eea
Therefore we get for non-flavour tagged sum 
\bea
\Gamma_1\left[ \int _0^{\infty}dt|{\cal A}(B_s   \to f;t)|^2  + \int _0^{\infty}dt|{\cal A}(\bar B_s   \to {\bar f};t)|^2\right] = 
\nonumber \\  
= 2|A_f|^2 + 2|\bar A_{\bar f}|^2 + 2y_s
\left(  ( |A_f|^2 + |\bar A_{\bar f}|^2)  -2 {\rm Re}\left( \frac{q_s}{p_s}\bar A_{\bar f} \otimes A^*_f   \right)    \right)
\label{BSNOPROD}
\eea
If there is a production asymmetry one gets: 
\bea
\Gamma_1\left[ \int _0^{\infty}dt|{\cal A}(B_s   \to f;t)|^2  + \int _0^{\infty}dt|{\cal A}(\bar B_s   \to {\bar f};t)|^2\right] = 
\nonumber \\  
=  2|A_f|^2 + 2|\bar A_{\bar f}|^2 +2y_s
\left(  ( |A_f|^2 + |\bar A_{\bar f}|^2)  -2 {\rm Re}\left( \frac{q_s}{p_s}\bar A_{\bar f} \otimes A^*_f  \right)     \right) 
+ \frac{2(1-2\rho (B_s))}{x_s}{\rm Im}\left( \frac{q}{p} \bar A_{\bar f} \otimes A^*_f\right) 
\label{BSPROD}
\eea
While the strength of indirect CP violation has not measured in time-resolved data on $B_s \to \psi \phi$, 
there are some evidence that it might be significantly larger than the CKM prediction of around 
sin$2\beta_s \sim 0.03 - 0.05$. That situation should be more clarified in one to three years. The $2y_s$ term 
can give us useful information about the dynamics of $B$ decays, but itself does not represent a CP 
asymmetry. 
In principal if there is a production asymmetry -- it could be tracked by the Cabibbo suppressed transition
$\bar B_s \to \psi K^+\pi^-$ vs. $B_s \to \psi K^-\pi^+$ -- one could obtain 
Im$\left( \frac{q_s}{p_s}\bar A_{\bar f} \otimes A^*_f   \right)$; however it is greatly suppressed in 
$B_s \to K_S\pi^+\pi^-$, $K_SK^+K^-$ by $1/x_s$ -- i.e., the spectacularly fast oscillation.

%%%%%%%%%%%%%%
\subsubsection{Comments on $B^0$ transitions}
%%%%%%%%%%%

Our goal for $B^0$ transitions is to analyses the impact of ND on {\em direct} CP asymmetries in 
{\em three-body} decays in CKM suppressed channels. 

Indirect CP violation affects all transitions of a {\em given} meson -- $B_d$, $B_s$ and $D^0$ -- in the same way. 
For $B_d$ we have measured it with good accuracy in $B_d \to \psi K_S$ with CKM dynamics as the leading source. 
The SM prediction tell us that $|q_d/p_d|$ can differ from unity by less than 0.001. 

For $B_s$ transitions some evidence has been found in $B_s \to \psi \phi$ and $B_s \to l^-X$ processes for a large 
impact of ND. We expect that evidence will be validated or reject with good accuracy in the foreseeable future from 
LHCb, CMS and ATLAS.  For the time being one can use two scenarios, namely 
\begin{itemize}
\item 
Case CKM: sin$2\beta_s \sim 0.03 - 0.05$, $\left| |q_s/p_s| - 1\right| < 0.0001$; 
\item 
Case CKM + ND: sin$2\beta_s \simeq 0.11 \pm 0.02$, $\left| |q_s/p_s| - 1\right| \simeq 0.003$ 
\end{itemize}
keeping in mind that such cases will be decided about future data on $B_s \to \psi \phi$ and 
$B_s \to l^-DX$. We consider the impact of ND in $B_s \to K_SK^+K^-$, $K_S\pi^+\pi^-$.

%%%%%%%%%%%%%
\subsubsection{$D^0\to h^+h^-h^0$}
\label{THREEDTOY}
%%%%%%%%%%%%

For $D^0$ transitions both $x_D$ and $y_D$ are small and probably of similar size, but 
$|q_D/p_D|$ could differ from unity by up to 30 \%.   
Integrating over all times $t$ of $D^0$ decays one gets
\bea
\int _0^{\infty}dt|{\cal A}(D^0   \to f;t)|^2 + \int _0^{\infty}dt|{\cal A}(\bar D^0   \to \bar f;t)|^2 \propto 
\nonumber \\ 
|A_f|^2 + |\bar A_{\bar f}|^2 - y_D {\rm Re} \left( \frac{q}{p} + \frac{p^*}{q^*}  \right)\bar A_{\bar f} \otimes A^*_f 
+ x_D {\rm Im} \left( \frac{q}{p} - \frac{p^*}{q^*}  \right)\bar A_{\bar f} \otimes A^*_f 
\eea
An production asymmetry leads to
\bea
\rho (D^0) \int _0^{\infty}dt|{\cal A}(D^0   \to f;t)|^2 + (1- \rho(D^0))\int _0^{\infty}dt|{\cal A}(\bar D^0   \to \bar f;t)|^2 \propto &&
\nonumber \\ 
\rho (D^0)|A_f|^2 +(1- \rho (D^0)) |\bar A_{\bar f}|^2 
- y_D {\rm Re}\left[  \left( \rho (D^0)\frac{q}{p} + (1 - \rho (D^0))\frac{p^*}{q^*}  \right)\bar A_{\bar f} \otimes A^*_f\right] + 
\nonumber \\
+ x_D {\rm Im} \left[ \left( \rho (D^0)\frac{q}{p} - (1- \rho (D^0))\frac{p^*}{q^*}  \right)\bar A_{\bar f} \otimes A^*_f \right] 
\eea
In $D^0$ decays the interplay of indirect and direct CP violations is not so clear mostly due to very slow 
oscillation. Therefore we will consider scenarios with $\left| |q_D/p_D| - 1\right| \simeq 0.1, 0.03$,  
$|A_f|/|\bar A_{\bar f}| \simeq 0.1, 0.03$ and the relative phase of $q_D/p_D$ and $\bar A_{\bar f} \otimes A^*_f$. 
We will present studies in a future paper. 

%%%%%%%%%%%%%%%%
\subsubsection{Comments on ND scenarios}
%%%%%%%%%%%%

For analyzing CP asymmetries in $h^+h^-h^0$ final states one has to include not only $PV$ final states, but 
also $SP$ final states like scalar $\sigma$ and $\kappa$ resonances. One reason for that is that exchanges from 
charged Higgs fields will introduces CP asymmetries already on the zero-loop processes, and they will affect 
scalar resonances more than $PV$ final states.

%%%%%%%%%%%%%%%%
\subsection{Future Progresses in Describing Dalitz Plots}
\label{FUTTH}
%%%%%%%%%%

When LHCb and the B factories at SLAC and KEK was planned and approved CKM theory had a 
competition with other models for the leading source of CP violation in heavy flavour transitions. 
BaBar and Belle have found with great success that CKM provides at least the leading source of the 
establish CP asymmetries in $B_d$ decays.  LHCb is in a great position to find whether CKM is also the leading source 
of CP asymmetries in $B_s$ decays even for $B_s \to \psi \phi$ transitions that are largely reduced in CKM. 

In addition LHCb and Super-Flavour Factories have to deal with the difficult task to find {\em non}-leading source(s) 
of CP asymmetries in suppressed decays. There are several candidates for that ND -- even there is no 
`standard' version of SUSY, let alone for other NDs. 

The `Miranda Procedure'  can allow us to find a clean evidence for a CP asymmetry without a 
theoretical input. It will encourage much more theoretical progress on our 
understanding on soft QCD, which can make use of other theory tools obtained from hadronic dynamics. 
The tasks one faces in $B$ and $D$ decays into three-body final states are not quite as challenging as for the 
astronomers mentioned above:  we know the locations where clear CP asymmetries can occur in 
$B^0$ and $D^0$ transitions -- in particular in $K_S\pi^+\pi^-$ and $K_SK^+K^-$ final states: They get 
sizable contributions from $K_S\rho^0$, $K^{*\pm}\pi^{\mp}$ and $K_S\phi$. it has been shown that the 
Breit-Wigner parameterization provides a good approximation for vector mesons 
like $\rho$, $K^*$ and $\phi$. However scalar resonances will in general not be described that way; 
for a final state $K_S f_0(980)$ it might give a decent description due to its relatively narrow width, but {\em not}  
for $K_S\sigma (600)$ or $\kappa \pi$ due to their wide widths. Still using a Breit-Wigner parametrization 
published data include true scalar resonances under `non-resonance' label; more theoretical analyses including the 
treatment of chiral dynamics is needed, very topical -- and possible now based on progress in the last few years. 
We know that they have to be performed {\em separately} for different $B/D \to 3h$ transitions. We should understand 
the following: if present data for $K\pi^+\pi^-$ final states are best fitted without any $\sigma \to \pi^+\pi^-$ 
contribution or for two different $\sigma_{1,2} \to \pi^+\pi^-$ one, one should not ignore one with the usual 
single $\sigma (600)$ as long as it gives a satisfactory description. One needs more experimental and 
theoretical analysis. 

There are several important reasons to analyze the production of scalar resonances in the detailed way. Let us 
just sketch one: Many models of ND contain physical charged Higgs states that can introduce CP asymmetries 
even through their {\em tree-level} exchanges. Obviously scalar Higgs exchanges will leave their `footprint' in the 
production of scalar resonances with more weight than for pseudoscalar and vector states. Therefore 
such ND will produce more `readable' impacts in CP asymmetries with scalar resonances and their interferences 
with pseudoscalar-vector final states. Therefore we can first focus on the known location of the peaks of the $\rho$, 
$K^*$ and the $f_0(980)$ and their widths from available data. Using this general input from theory we can 
generate binning for $B^0 \to K_Sh^+h^-$ and do it separately for $B_d$ and $B_s$ transitions.

%%%%%%%%%%%%%%
\section{Second Generation `Miranda Procedure'}
%%%%%%%%%%%
\label{2MIRPRO}

The procedure given in Ref. \cite{MIRANDA1} is obviously powerful for finding CP asymmetries and even 
`localizing' them. However one wants to make it more quantitatively and to understand its source(s) --  
in particular for $B$ transitions CKM gives sizable `backgrounds' when searching for ND. 

The goal is to find a way to evaluate the strength of {\em local} effects, namely to have numbers that
are equivalent to the asymmetry between (time) integrated rates, $A_{\mathrm{CP}}$.
The key idea is to divide the combined Dalitz plot into bins with equal populations. If one knows
the number of bins where CP is violated, the one can compute a local average value of
$A_{\mathrm{CP}}(s_1,s_2)$, since the number of events is proportional to the number of bins.

Each bin has $N=N^+ + N^-$ events, with $N^+$ and $N^-$ being the numbers of $B$ and
$\bar B$ candidates. We assume that there are regions in the Dalitz plot with at least
a few tens of bins in which positive events ($N^+$) occur with the same probability $p$.  
$N^+$ follows a binomial distribution with expected value and variance given by
\beq
E[N^+] = Np, \hskip .5cm V[N^+]=Np(1-p)
\eeq
When $N$ is large enough (at least a few tens of events), the Central Limit Theorem ensures that $N^+$
follows a normal distribution, allowing one to write exact expressions for the expected difference
$N^+ - N^-$. In this case one has
\beq
A_\mathrm{{CP}}^{\mathrm{bin}} = \frac{N^+ - N^-}{N} = \frac{2N^+}{N}-1
\eeq
with
\beq
\mu = E[A_\mathrm{{CP}}^{\mathrm{bin}}] = \frac{2E[N^+]}{N} - 1 = 2p - 1
\label{mean}
\eeq
and
\beq
\sigma^2= V[A_\mathrm{{CP}}^{\mathrm{bin}}] = \frac{4V[N^+]}{N^2} = \frac{4p(1 - p)}{N}
\label{sigma}
\eeq

If CP symmetry is conserved -- hereafter we assume that there is no other source of charge asymmetry -- 
the probabilities of positive and negative events are equal, $p=1/2$. One therefore has
\beq
\mu = 0 , \hskip .5cm \sigma^2=\frac{1}{N}
\eeq

When CP is violated the Dalitz plot will have regions with and with{\em out} asymmetries. 
Therefore, the distribution of  $A_{\mathrm{CP}}^{\mathrm{bin}}$ will be
a superposition of a Gaussian with $\mu = 0$ and $\sigma=1/\sqrt{N}$ plus 
some other function representing the CP violating bins. The form of the latter depends on how 
CP violation occurs in the Dalitz plot and also on the specific final state.

Three-body final states result, in general, of a cascade process in which the heavy meson decays to a 
resonance plus a `bachelor' hadron. 
The decay amplitude of a heavy meson $P$ is usually modeled by 
a coherent sum of resonant amplitudes, weighted by constant complex coefficients 
\beq
\mathcal{M} = \sum_i c_i A_i, \hskip .5cm c_i = a_i e^{\delta_i}.
\eeq
CP violation results in  a difference between $\mathcal{M}(P)$ 
and $\overline{\mathcal{M}}=\mathcal{M}(\overline{P})$. More specifically, $\mathcal{M}$ and 
$\overline{\mathcal{M}}$ may  
\begin{itemize}
\item 
differ by the {\em magnitude} of a set of resonant modes 
\item
or a difference between their {\em relative phases} 
\item
or a {\em combination} of both. 
\end{itemize}
Rescattering at the hadronic level is a long distance effect
that mixes different final states, e.g. $\bar KK\pi \to \pi\pi\pi$, and this is another 
source of CP violation. 

In this paper, we consider three possibilities: (i) CP violation due to re-scattering 
as a constant excess of one specie over the other limited to some region of the Dalitz plot;
(ii) CP violation through a difference in the magnitude of a resonant amplitude; 
(iii) a difference between relative phases.

The case of constant CP violation is the simplest: it leads to asymmetries 
that have always the same sign. An uniform and localized excess of one charge state over the other is
equivalent to a constant value of $p$. In this case the
distribution of CP violating bins will be a Gaussian with mean and sigma given by Eqs.(\ref{mean})
and (\ref{sigma}). Integration over the phase space results is an observable global $A_{\mathrm{CP}}$.

Differences in magnitude would also correspond, in principle, to a constant $p$. However, the net effect 
depends on the resonance spin and on the contribution of the other resonances. Angular momentum
conservation constrains the angular distribution of the decay products. For vector particles, for example, 
the Breit-Wigner is modulated by a spin amplitude which is proportional to the cosine of an helicity angle.
In the region where the momentum configuration is such that the helicity angle is 90$^{\circ}$, the
amplitude goes to zero. The relative contribution of the CP violating amplitude varies from bin to bin,
in spite of the constant difference in its magnitude. Even in the case of a scalar resonance (constant spin
amplitude), one needs to take into account the contribution from other resonant amplitudes to the CP 
violating bins, which in general is {\em not} constant. Also in this case an integration over the phase space 
results in an observable global $A_{\mathrm{CP}}$, 
although the local effect will be always diluted by the relative contribution of the CP violating 
amplitudes. 

Differences in phases are the most complex case. As illustrated in \cite{MIRANDA1}, such differences  
lead to asymmetries that change sign across the Dalitz plot. Integration over the phase space could
result in a null asymmetry, in spite of large local effects. In the simplest case of two
resonances, Eq.(\ref{acp}) would read
\beq
A_{\mathrm{CP}}(s_1,s_2) = 
\frac{2\sin(\Delta\phi^W) \sin(\Delta\delta(s_1,s_2)) |\mathcal{A}_2\mathcal{A}_1|}
{1 + |\mathcal{A}_2\mathcal{A}_1|^2 + 2|\mathcal{A}_2\mathcal{A}_1|\cos(\Delta\phi^W) 
\cos(\Delta\delta(s_1,s_2))}
\label{acpl}
\eeq
The asymmetry  is driven by $\Delta\delta(s_1,s_2)$ due to 
interfering Breit-Wigner functions spread over the phase space. This
is equivalent to having a different value of $p$ for each bin. The distribution of
$A_{\mathrm{CP}}^{\mathrm{bin}}$ for the CP violating bins therefore depends strongly on the
final state, on which resonances and with which relative phases it is built of. 

An important effect is the charge asymmetry induced by different production mechanisms. This is
not possible in $p\bar p$ collider, but it may occur in asymmetric collisions (like for LHCb data). The production asymmetry may as large as a 1-2\% effect. Since it depends on the heavy 
meson momentum, it may vary across the Dalitz plot. In the following examples we assume that any eventual
production asymmetry would lead only to a global effect, constant throughout the Dalitz plot.

%%%%%%%%%%%%%%%%
\subsection{Comment on CPT Constraints} 
\label{CPTCON}
%%%%%%%%%%%

It is mentioned usually that CPT symmetry gives equality of masses and total widths of $P$ and $\bar P$. 
However it gives also equality of {\em different} classes of final states, where mixing happens; 
some general comments are given in Sect. 4.10 in \cite{CPBOOK}.   
Up to isospin violation one has for example: 
\bea
\Gamma (B_{u,d,s} \to 2 \pi, \, K\bar K, \, 4 \pi, \, 2K2\bar K, \, 6 \pi) &=& 
\Gamma (\bar B_{u,d,s} \to 2 \pi, \, K\bar K, \, 4 \pi, \, 2K2\bar K, \, 6 \pi) \\
\Gamma (D_{u,d,s} \to 2 \pi, \, K\bar K, \, 4 \pi) &=& 
\Gamma (\bar D_{u,d,s} \to 2 \pi, \, K\bar K, \, 4 \pi)
\eea
While mixing happens -- and diagrams show it -- we have little {\em quantitive} control over it. 
In a qualitative way one expects correlations like between CP asymmetries in 
$D^0 \to K^+K^-$ and $D^0 \to \pi^+\pi^-$ or in $\bar B_d \to K^-\pi^+$ and $\bar B_d \to K_S\pi^0$ 
etc. As emphasized before that CP asymmetries with three-body final states will give us more information 
-- and probably crucial one --  
about the underlying dynamics. Since SM produce sizable CP violation in $b \to s q\bar q$ with 
$q = u, d, s$ one expects sizable CP asymmetries in 
$B_d \to K_S \rho^0/K_S \sigma/K^+ \rho^-/\kappa \pi$ and higher resonances etc. with 
{\em different} signs compensate for CPT relation -- but only qualitatively in practice.

%%%%%%%%%%%%%%%%%%
\subsection{`Miranda Procedure' for $B_d$ Three-Body Decays}
\label{REALBD}
%%%%%%%%%
 
We give `realistic' studies for $B_d\to K_S\pi^+\pi^-$, where we have decent data and some
information about the resonant structure \cite{babar,belle}. In all studies we consider time integrated,
tagged samples. In each exercise two samples of $B^0,\overline{B^0} \to K_S\pi^+\pi^-$ were 
simulated independently using the same set of resonant amplitudes, namely
$K_S \rho, K_S f_0(980), K_S f_0(1370)$, $K^*(892)\pi$ and $K_S \chi_c$. The samples are generated with
CP violation seeded in three different ways, as described above.

A few remarks are in order:
\begin{itemize}
\item 
Indirect CP violation has been very well measured in $B_d \to \psi K_S$ with 
sin$2\phi_1/\beta = 0.658 \pm 0.024$; this observable enters in many transitions as an input quantity. 
However for the time integrated $B_d$ + $\bar B_d$ rates indirect CP asymmetry cannot contribute -- 
{\em unless} there is a production asymmetry for $B_d$ vs. $\bar B_d$.  
\item 
Direct CP violation can occur even in the time integrated $B_d$ + $\bar B_d$ rates. 
\item 
In the SM one has three quark-level processes, namely two tree-level $b \to u \bar u s$ and $b\to s \bar u u$, 
where the second one is generated by QCD radiative corrections, and the loop Penguin $b\to s+g's$. 
They produce another $\bar dd$ and $\bar uu$ pair for the final state $K_S\pi^+\pi^-$ and a $\bar ss$ 
for $B_d \to K_SK^+K^-$. The Penguin operator $b\to s +g's$ generates no weak phase; since it produces 
a $\Delta I=0$ transition, there is no appreciable relative strong phase from this contribution. On the 
other hand $b \to u \bar u s$ and $b\to s \bar u u$ represent a combination of $\Delta I = 0$ and 
$\Delta I=1$ amplitudes that in general will have different strong phases. 
As an example for ND: Charged Higgs exchanges would probably affect mostly $b\to u \bar u s$ and 
$b\to s \bar u u$, introduce another weak phase and different strong phases. Furthermore they should 
affect final states with 
pseudoscalar-scalar more than pseudoscalar-vector. The impact of ND in direct CP violation should 
be clearer in the former than the latter, since the latter `suffer' from a larger `background' from CKM.   

\item 
Sizable contributions from final states $K_S\rho$, $K^*(892)\pi$, $K^*(1430)\pi$, $K_S f_0(980)$ and 
$K_S f_0(1500)$ have been reported. No obvious contributions from $K_S\sigma$ and/or $\kappa \pi$ have been found, 
but might be hidden under the `no-resonance' listing without 30 \% of the rate of $B_d \to K_S\pi^+\pi^-$. 
There are several theoretical arguments that such final states $K_S\sigma$ and $\kappa \pi$ should 
occur in an appreciable way.     
\item 
While CKM dynamics has been found to produce at least the leading source of indirect CP violation in 
$B_d - \bar B_d$ oscillations, ND could still represent up to about 20 \% of it. While no clear deviation from 
CKM theory has been found in direct CP asymmetries in $B_d$ decays, ND could produce significant contributions. 
One expects that the weight of ND in direct CP asymmetries will change differently for classes of channels 
like pseudoscalar-vector vs. pseudoscalar-scalar.  

\end{itemize}

 %%%%%%%%
 \subsubsection{$B_d/\bar B_d \to K_S\pi^+\pi^-$  -- Constant CPV} 
 %%%%%%%%%%%%%%
 
Direct CP violation has been found in $B_d \to K^+\pi^-$ around 10 \%. 
No sign has been found in $C(B_d \to K^0\pi^0) = 0.00 \pm 0.13$. Yet one could 
find sizable impact with future data.  

The first and simplest study of the `Miranda Procedure'  
refers to the case where one has one single source of direct CP violation
acting on a given region of the Dalitz plot.  
The CP violation is seeded as a 10\% excess of $B^0$ over $\overline{B}^0$ in the region
$s_{K_S\pi^+},s_{\pi^+\pi^-}<7.5$ GeV$^2$/$c^4$. We have generated 300K $B^0$ and 
330K $\overline{B}^0$ decays dividing the combined Dalitz plot into 256 
bins of equal population. This excess of $B^0$ over $\overline{B}^0$ events is equivalent to a global
$A_{\mathrm{CP}}$ of 4.76\%. The distribution of the $A_{\mathrm{CP}}^{\mathrm{bin}}$
across the Dalitz plot is shown in Fig.1. 

%In Fig.2 we present the distribution of $A_{\mathrm{CP}}^{\mathrm{bin}}$.
Having only one source of CP violation (constant $p$), 
the values of $A_{\mathrm{CP}}^{\mathrm{bin}}$ for the bins in the region
where CP violation was seeded are the same within statistical fluctuations. 
We therefore expect the $A_{\mathrm{CP}}^{\mathrm{bin}}$ for the CP violating bins to be also
distributed as a Gaussian.
The distribution in Fig.2 is  fitted by two Gaussian functions. The one representing the
CP conserving bins has fixed mean ($\mu = 0$) and sigma ($\sigma = 1/\sqrt{N}$), whereas
the parameters defining the second Gaussian are free.

The average value of $A_{\mathrm{CP}}^{\mathrm{bin}}$ in the region where CP violation was seeded 
is the mean of the second Gaussian, (13.64 $\pm$ 0.25)\%.
The normalization of each Gaussian is the
number of bins that conserve/violate CP. There are 167$\pm$13 bins conserving CP and 89$\pm$9
bins in which CP is violated. The number of events is the same for all bins, so we can obtain the
global $A_{\mathrm{CP}}$ from the ratio of CP violating to the total number of bins, 
and from the average value of $A_{\mathrm{CP}}^{\mathrm{bin}}$,
\beq
A_{\mathrm{CP}} = \frac{n_2}{n_1+n_2} \ <A_{\mathrm{CP}}^{\mathrm{bin}}> = 4.98\pm0.54 \%
\eeq
We not only recover the global $A_{\mathrm{CP}}$ but also access the average 
$A_{\mathrm{CP}}^{\mathrm{bin}}$ and the fraction of events that violate CP and thus the 
localization of the source. 
 
This exercise clearly shows how the relatively large local effect is diluted when the CP violation 
strength is measured by the global $A_{\mathrm{CP}}$ (in this case,
by the ratio of the area of the CP violation region and the total Dalitz plot area).

\begin{figure}
\includegraphics[width=13cm]{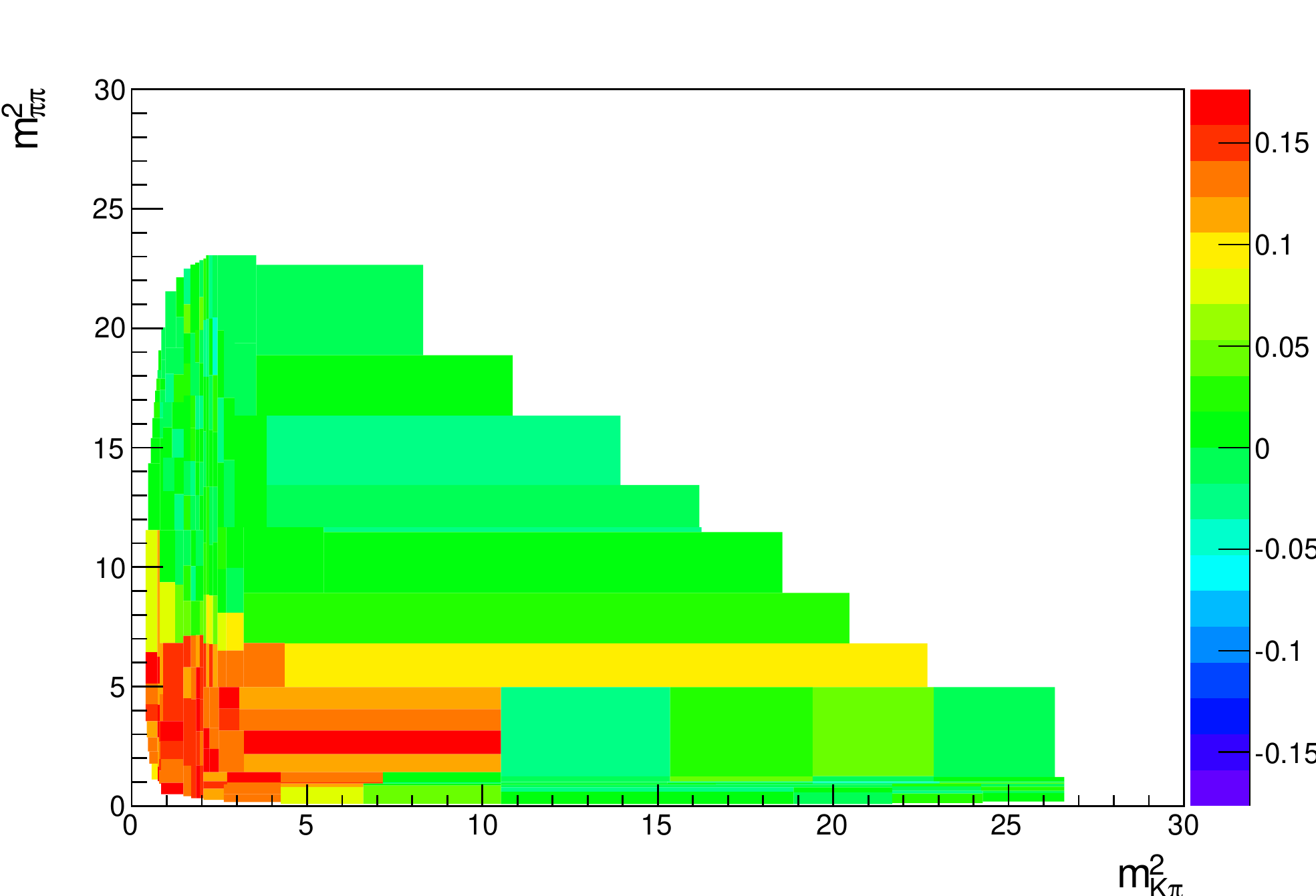}
\caption{Distribution of $A_{\mathrm{CP}}^{\mathrm{bin}}$ across the $B_d \to K_S\pi^+\pi^-$ Dalitz plot.
In this example a single source of CP violation -- constant excess of $B_d$ over $\bar B_d$ restricted to 
the low $K_S\pi^+/\pi^+\pi^-$ mass region -- was simulated. Bins have different size in order to
contain the same number of events.}
\end{figure}

\begin{figure}
\includegraphics[width=13cm]{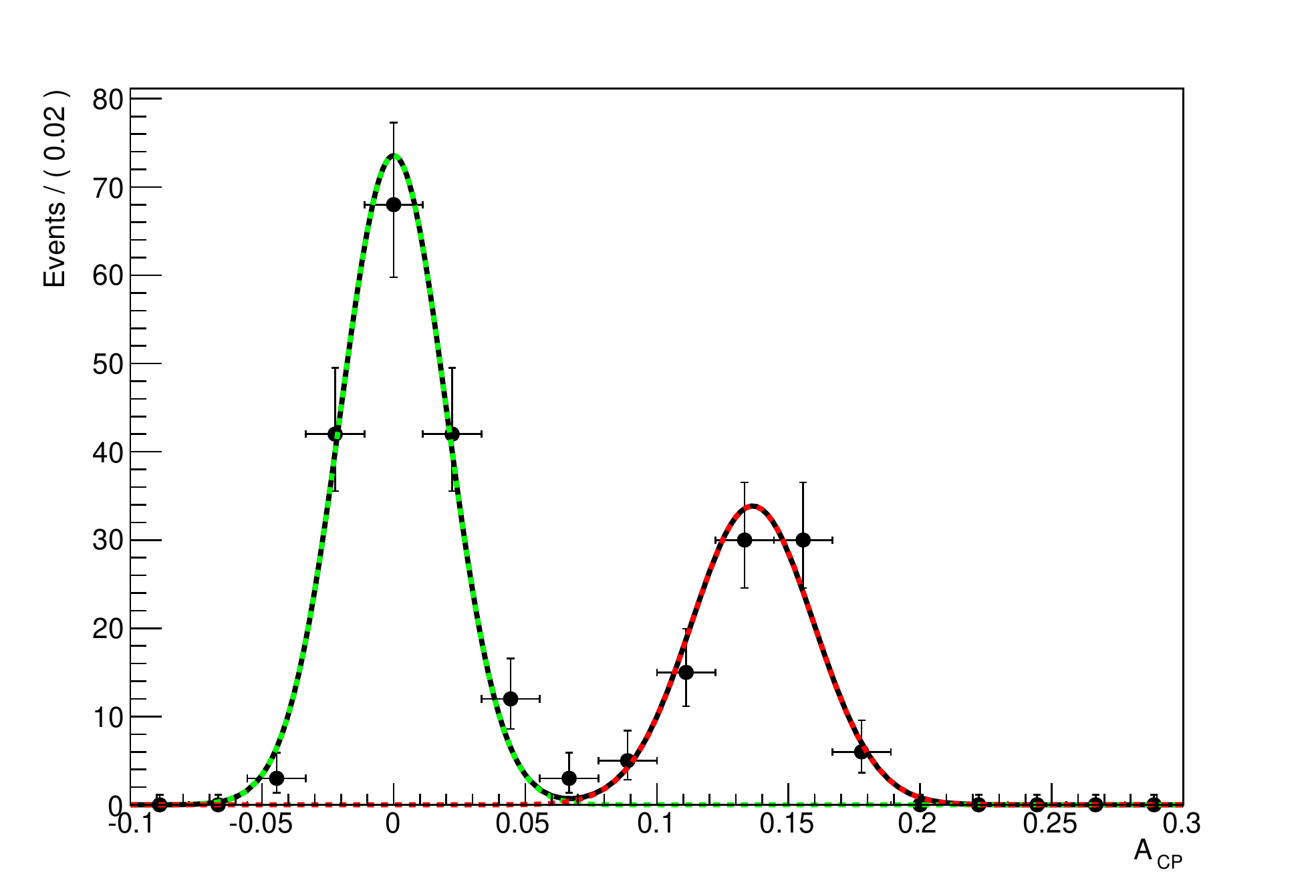}
\caption{Distribution of $A_{\mathrm{CP}}^{\mathrm{bin}}$ for  single, constant source of CP violation.
The distribution was fitted to two Gaussians. The one centered at zero represents bins where CP is conserved,
whereas the second Gaussian represents the CP violating bins.}
\end{figure}

%%%%%%%%%%%%%%%%%%
\subsubsection{ $B_d/\bar B_d \to K_S\pi^+\pi^-$ -- Difference in Magnitudes}
%%%%%%%%%%%%%%

In this example CP violation is seeded as a 10\% difference  in the magnitude 
of the resonant mode $K^*(892)\pi$, $a_{K^*(892)\pi} = 0.9 \overline{a}{K^*(892)\pi}$. 

The total decay rates of $B^0$ and $\overline{B^0}$ are 
proportional to the integral over the phase space of  $\mathcal{M}$ and $\overline{\mathcal{M}}$,
respectively. In the present example, this means a global CP asymmetry of 2.1\%.
Samples of  300K $B^0$ and 287K $\overline{B}^0$ decays were generated.
The combined $B^0$ and $\overline{B^0}$ Dalitz plot was divided into 1024 bins of equal population.

The extra $B^0$ events are distributed in the bins along the $K^*(892)$ band,
and the resulting $A_{\mathrm{CP}}^{\mathrm{bin}}$ across the Dalitz plot is shown in Fig.3; 
note that the values of $A_{\mathrm{CP}}^{\mathrm{bin}}$ in the CP violating region are always 
positive. 

In Fig.4 the distribution of $A_{\mathrm{CP}}^{\mathrm{bin}}$ for all bins is presented. 
The bins with no CP violation have equal number of
$B^0$ and $\overline{B^0}$ decays, within statistical fluctuations, resulting on a Gaussian 
distribution of $A_{\mathrm{CP}}^{\mathrm{bin}}$ with $\mu=0$ and $\sigma = 1/\sqrt{N}$. 
The distribution of $A_{\mathrm{CP}}^{\mathrm{bin}}$ for the CP violating bins is again parameterized 
by Gaussian, but this is used as an effective representation ($p$ is no longer constant).  
As in the previous example, the distribution 
in Fig.4 is fitted to two Gaussian functions, one with fixed mean and sigma representing the CP 
conserving bins.

\begin{figure}
\includegraphics[width=13cm]{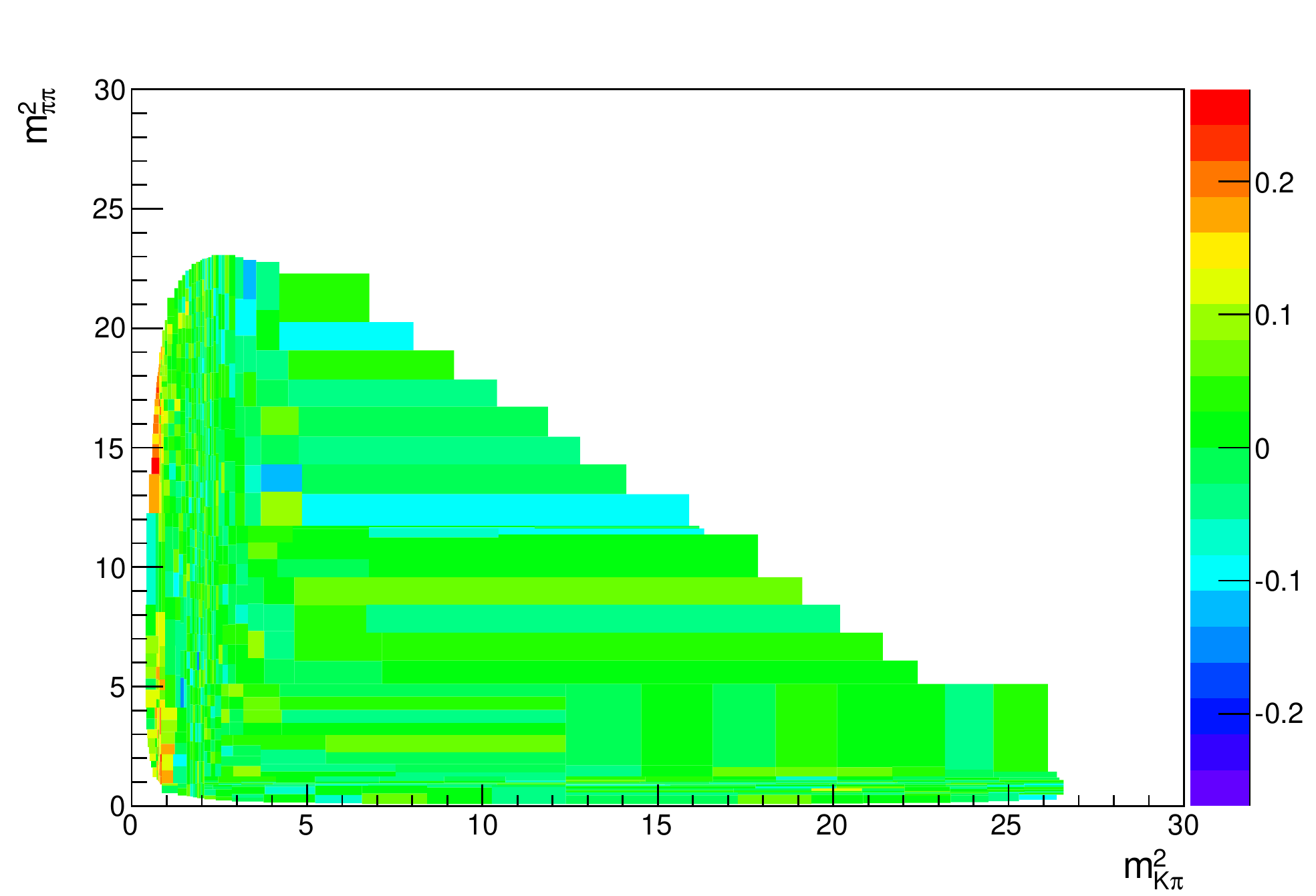}
\caption{Distribution of $A_{\mathrm{CP}}^{\mathrm{bin}}$ across the $B_d \to K_S\pi^+\pi^-$ Dalitz plot
for the case of CP violation seeded as a difference in the magnitude of the $\rho K_S$ mode between $B_d$ and
$\bar B_d$. The excess of $B_d$ over $\bar B_d$ events is concentrated along the $\rho$ band. Note that
the values of $A_{\mathrm{CP}}^{\mathrm{bin}}$ for these bins are always positive.}
\end{figure}

\begin{figure}
\includegraphics[width=13cm]{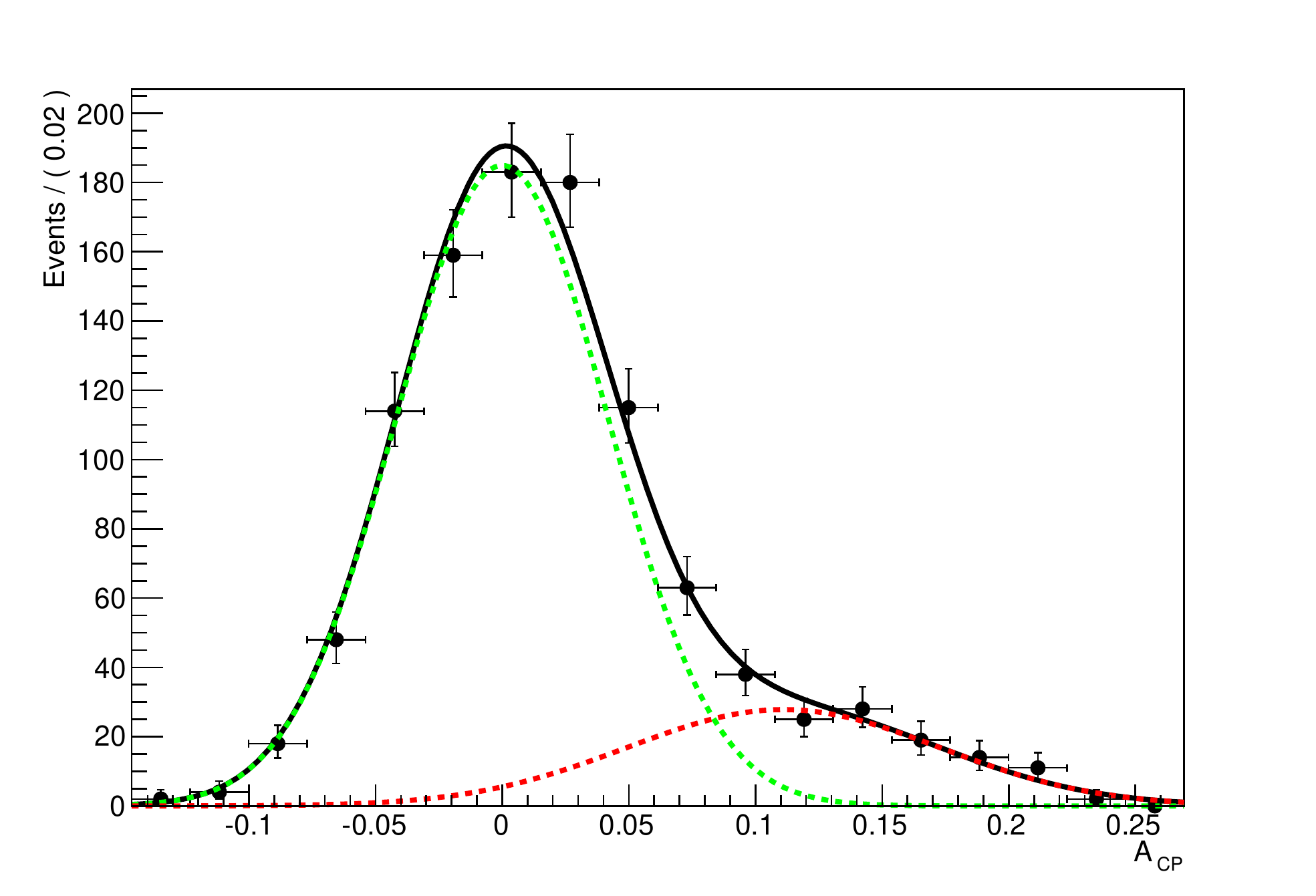}
\caption{Distribution of $A_{\mathrm{CP}}^{\mathrm{bin}}$ for the case of CP violation through
a  difference in the magnitude of the $\rho K_S$ mode. The  Gaussian in red is an empirical representation of 
$A_{\mathrm{CP}}^{\mathrm{bin}}$ for the CP violating bins.}
\end{figure}

We find 838 $\pm$ 46 bins with no CP violation and
186 $\pm$ 38 bins with average value of $A_{\mathrm{CP}}^{\mathrm{bin}}$ (11.1 $\pm$ 1.7)\%. 
From these parameters we extract the global asymmetry, $A_{\mathrm{CP}}=2.1 \pm 0.1$\%.

These exercises show that the observable $A_{\mathrm{CP}}^{\mathrm{bin}}$ carries the relevant 
information about local asymmetry. In both cases the  CP violation is restricted to certain regions
of the Dalitz plot, but leads to a global asymmetry. The local effects are much more intense than
the phase space integrated ones. We showed that the later can be recovered in a consistent way.

%%%%%%%%%%%%%%%%%%%%%%%%%%
\subsubsection{ $B_d/\bar B_d \to K_S\pi^+\pi^-$ -- Difference in Relative Phases}
%%%%%%%%%%%%%%%%%%%%%%%

We now discuss a more general case, where the CP violation occur via a difference 
between $B$ and $\bar B$ in relative phase of a given set of resonances.
This is a much more difficult and subtle situation, which depends strongly
on the final state characteristics: (i) Which resonances are present. 
(ii) What are their relative phases. 
(iii) Is there an sizable contribution from scalars \cite{JOBS}. 

Two independent samples were generated using the same set of resonances as in the previous examples.
A 60$^{\circ}$ phase difference in the $\rho K_S$ mode was introduced between $B_d$ and $\bar B_d$.
The combined Dalitz plot was then divided into 1024 bins. The seeded phase difference is
large, causing local asymmetries that can be as large as 80\%, shown in Fig.5. The {\em global} 
asymmetry, however, is small:  1.0\%. Due to the phase variation of the Breit-Wigner curve, the CP asymmetry
change sign along the $\rho$ band. In this case the integration over the phase space -- necessary to 
compute the total rates -- cancels out most of the effect of CP violation. This cancellation is 
clearly seen in Fig.6, which has an enlarged view of the Dalitz plot region where CP violation occur.

The interference between $\rho K_S$ and the other resonant modes, which is governed by the combined
strong phases of the Breit-Wigner curve, $\Delta\delta(s_1,s_2)$,
causes each bin to act 
as an independent source of CP violation. When one has repeated the same experiment many times,  
the values of $A_{\mathrm{CP}}^{\mathrm{bin}}$ for each bin would be distributed with a mean and
sigma given by Eqs.(\ref{mean},\ref{sigma}), respectively, each bin having its own value of $p$. 
The distribution of $A_{\mathrm{CP}}^{\mathrm{bin}}$ for all bins will have two components,
as in the other examples. The distribution from the CP violating bins would no longer be a Gaussian,
but some function that is particular to each specific final state.

In general there would be as many bins with {\em positive} and {\em negative} 
$A_{\mathrm{CP\pm }}^{\mathrm{bin}}$,
so the procedure adopted in the previous examples would underestimate the measurement of
the average asymmetry in this case. As before, we can fit the distribution to a Gaussian for the CP 
conserving bins -- $\mu=0$ and $\sigma=1/\sqrt{N}$, but with unknown area -- plus one function for 
the CP violating bins. Having defined the Gaussian CP conserving bins, this can then be subtracted 
off,and two numbers could be computed: the average value of 
$A_{\mathrm{CP\pm }}^{\mathrm{bin}}$ for
the regions where the asymmetry is either positive or negative.
 
\begin{figure}
\includegraphics[width=13cm]{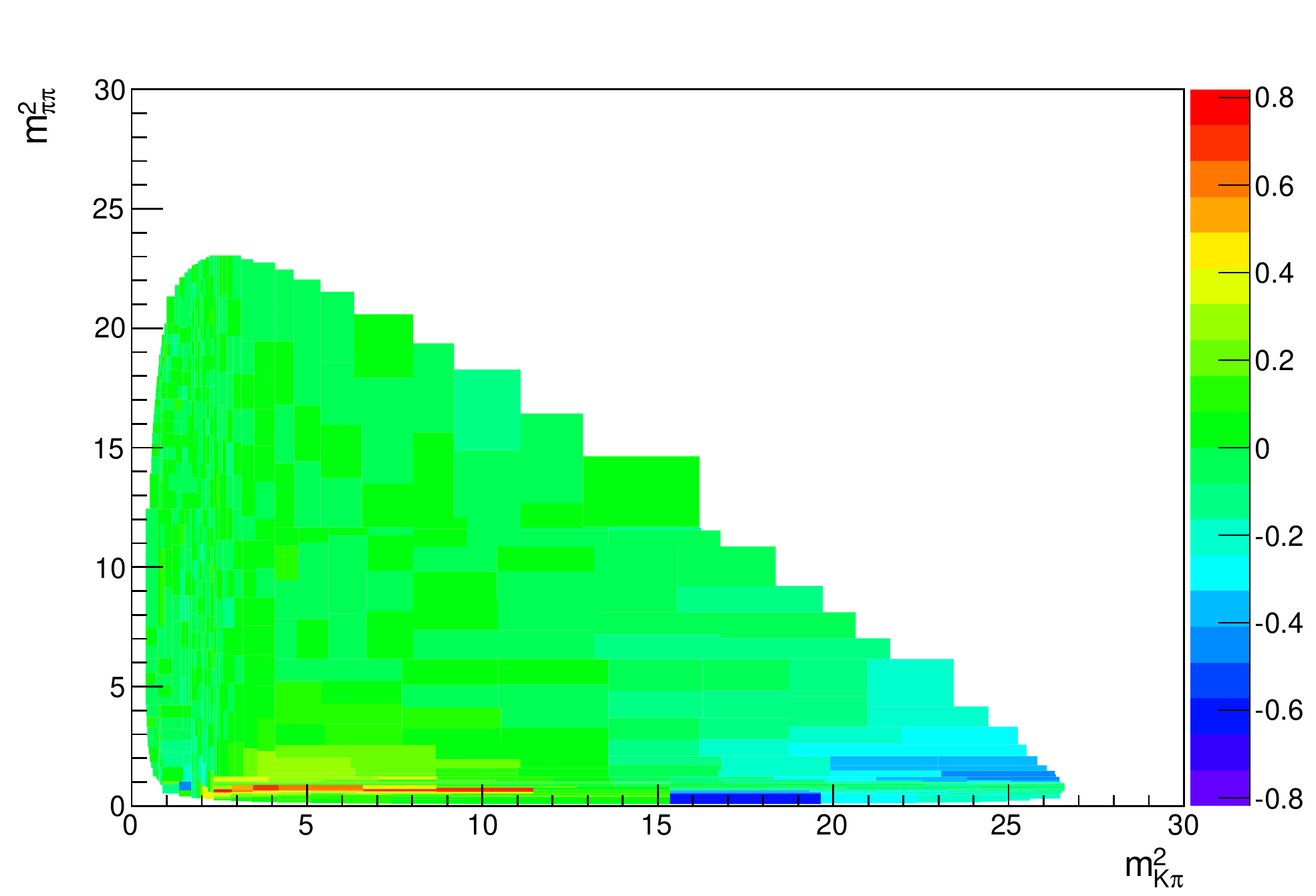}
\caption{Distribution of $A_{\mathrm{CP}}^{\mathrm{bin}}$ across the $B_d \to K_S\pi^+\pi^-$ Dalitz plot
for the third example. When the CP violation is seeded as a relative phase difference, the values
of $A_{\mathrm{CP}}^{\mathrm{bin}}$ change sign.}
\end{figure}

\begin{figure}
\includegraphics[width=13cm]{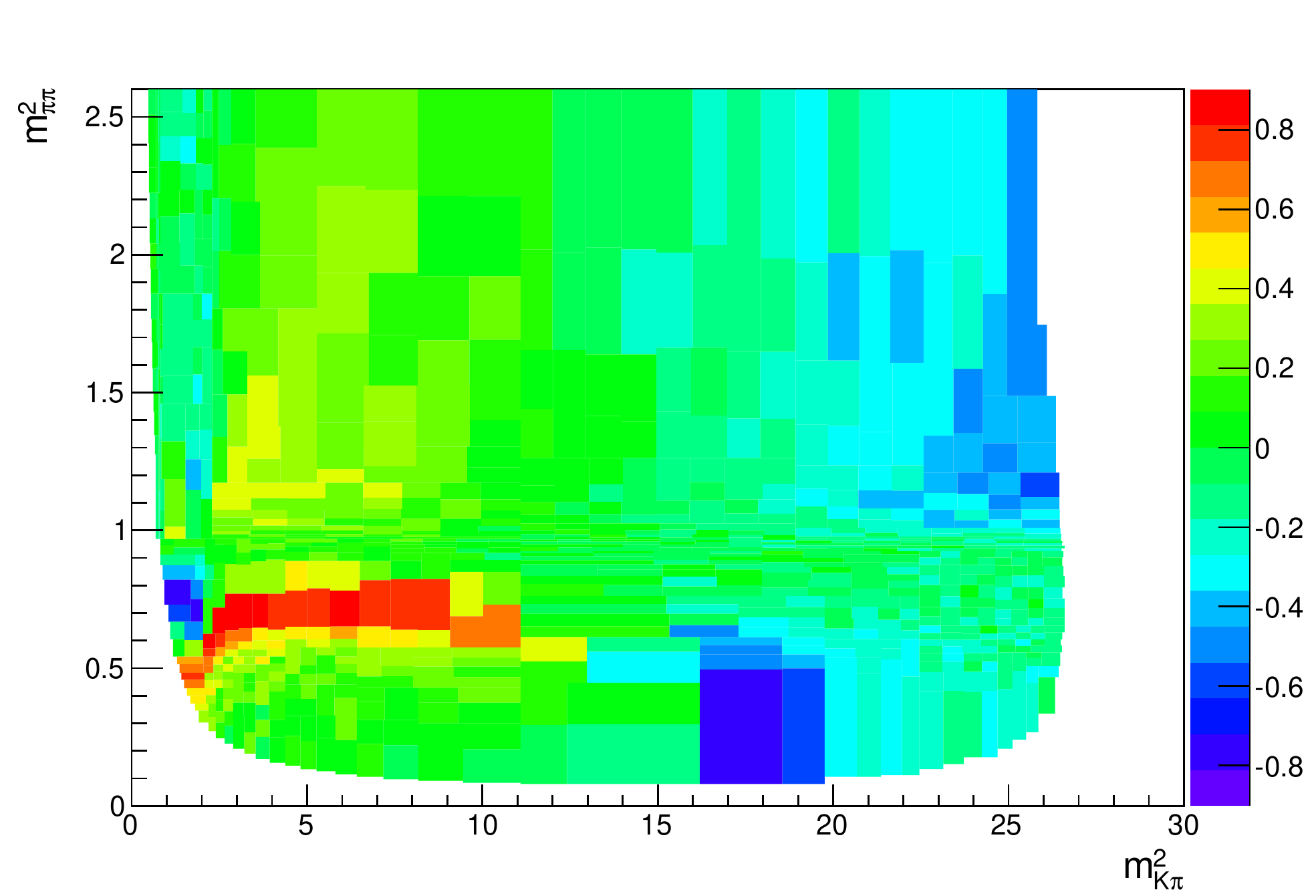}
\caption{Enlarged view of the CP violating bins of Fig.5.}
\end{figure}

\begin{figure}
\includegraphics[width=13cm]{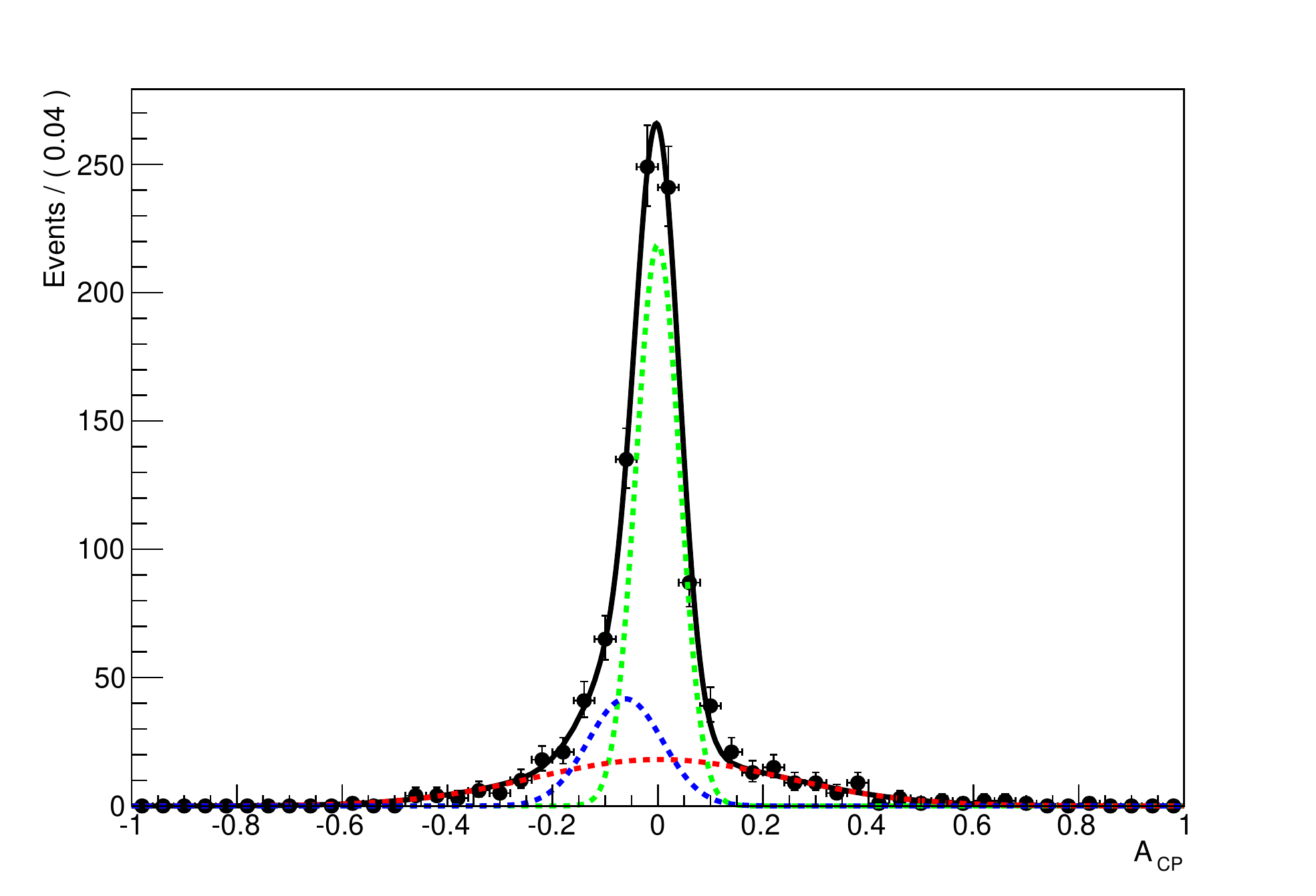}
\caption{Distribution of $A_{\mathrm{CP}}^{\mathrm{bin}}$ for the third example. In addition to the
Gaussian representing the CP conserving bins (green curve), two other Gaussians were used to empirically
represent the $A_{\mathrm{CP}}^{\mathrm{bin}}$ distribution of the CP violating bins (red and blue curves).}
\end{figure}
We illustrate this procedure in Fig.7, where the $A_{\mathrm{CP}}^{\mathrm{bin}}$ distribution
for the CP violating bins was empirically fit to two Gaussians. 
The fit yields 568 $\pm$ 83 bins in the CP conserving Gaussian,
and 466 $\pm$ 41 bins in which CP is violated. We then compute the weighted average value
of $A_{\mathrm{CP\mp }}^{\mathrm{bin}}$ for the negative and for the positive part of the 
distribution in Fig.7. This yields
\beq
<A_{\mathrm{CP-}}^{\mathrm{bin}}> = -(14 \pm 2)\% ,
\eeq
and
\beq
<A_{\mathrm{CP+}}^{\mathrm{bin}}> = (16 \pm 2)\%. 
\eeq
In order to test this procedure, we go to the limit of very high statistics. Since the
width of the distribution of CP conserving bins is $\sigma = 1/\sqrt{N}$, when $N$ is very 
large, the Gaussian gets very narrow, in practice restricted to the central bin of Fig.7.
We can then compute the weighted average in the negative and positive regions separately 
with a simple counting procedure, discarding the central bin.
The average values of  $A_{\mathrm{CP \mp }}^{\mathrm{bin}}$ obtained are
\beq
<A_{\mathrm{CP-}}^{\mathrm{bin}}> = -15 \% 
\eeq
and
\beq
<A_{\mathrm{CP+}}^{\mathrm{bin}}> = 22.2\% 
\eeq
in good agreement with the fitting procedure used in the more realistic scenario.

%%%%%%%%%%%%%%%%%%
\section{$B_s$ Three-Body Decays}
\label{REALBS}
%%%%%%%%%%%

At present, the experimental situation is very different for $B_s$ transitions even beyond the 
fact that $B_s - \bar B_s$ oscillations are very fast. 
\begin{itemize}
\item 
Within SM indirect CP violation is small -- i.e. sin$2\beta_s \sim 0.03 - 0.05$. Finding it significantly larger 
is a clear manifestation of ND. There is some evidence that indirect CP violation is larger than predicted by CKM; 
studies of $B_s \to \psi \phi$ in LHC data should clarify this issue and allow sin$2\beta_s$ as an input for 
searching manifestations of ND. 
\item 
If one indeed finds that CKM theory does not produce the leading source of indirect CP violation, there is a good 
chance that ND generates a larger contribution also to direct CP violations. 
\item 
Measuring $y_s$ more accurately will help in cross checking finding CP asymmetries in the sum of time integrated 
$B_s$ and $\bar B_s$ rates. 
\item  
There are no data for $B_s \to K_S\pi^+\pi^-$ or $B_s \to K_SK^+K^-$. 
One expects the Dalitz plots for these $B_s$ transitions very different for these $B_d$ transitions.  
\item 
The tree diagram $b\to u \bar u d$ and the Cabibbo disfavoured penguin 
one-loop diagram $b \to d +g's$ to generate direct CP violation in both CKM and ND. 
Again final states like $K_S \sigma$ should show clearly manifestations of the impact of ND. 

\end{itemize}

%%%%%%%%%%%%%%
\section{Outlook}
\label{CON}
%%%%%%%%

Present data from Belle, BaBar, CDF and LHCb and future ones from LHCb, Super-Belle and 
Super-BaBar have reached the status to probe the possible impact from ND in $B_{u,d,s}$ and 
$D_{u,d,s}$ with {\em accuracy} and {\em correlations}. Analyzing non-leptonic three-body final states 
there needs significantly more experimental efforts through `Miranda Procedure' -- but it will be 
awarded with more lessons about the underlying dynamics and deep insights into its `shape'. 

The {\em Miranda Procedure I} is a good way to show whether or not there is CP asymmetry
in three-body decays of $D$ and $B$ mesons. It can also tell us where in the Dalitz plot CP violation 
occurs and give hints of the kind of operators that are involved. A further development of this technique, 
presented here, is a necessary step towards a quantitative output. One should keep in mind, however,
the crucial difference between two- and three-body decays: while in former case CP asymmetries are
observed in total decay rates, in the latter case there are several options for CP violation manifestations.
CP asymmetries through phase difference
-- `favorited' by model builders -- are intrinsically complicated because each bin acts
as an independent source of CPV. Moreover, strong interactions governing phases across the Dalitz plot
are still out of control {\em quantitively}.
Accurate data on three-body final states will help efforts from theorists working on
HEP and Hadrodynamics/MEP. 

One should not forget about constraints from CPT symmetry, but 
those are not of {\em quantitative} value on a practical level; it should tell us to think about 
other channels in a qualitative level; theoretical inputs help here.  

The `Miranda Procedure'  helps greatly to `localize' CP asymmetries and find evidence for the impact of 
ND and its `shape'. It does not mean that theoretical inputs are not needed, but to focus on them. 
It should enhance interests from theorists working in HEP and HP/MEP. 

Applying `Second Generation of Miranda Procedure' is now at the `starting line' -- the `race' will 
proceed over many longer  `distances' with simulations and -- most importantly -- with real data: 
\begin{itemize}
\item
Time integrated and non-flavour tagged rates for $B_{u,s}/D_{u,d,s}$ decays; 
\item 
Flavour tagged ones for $B_{u,d,s}/D_{u,d,s}$; 
\item 
partially time integrated ones;   

\item 
$\tau \to \nu [K\pi/K2\pi/3K]$ decays. 

\end{itemize}   
One needs no more hardware -- `only' thinking and working.

\vspace{0.5cm}

{\bf Acknowledgments:} This work was supported by the NSF under the grant number PHY-0807959 and
by CNPq.

\vspace{4mm}

%%%%%%%%%%%%%%%%


\begin{thebibliography}{99}
\bibitem{JOBS}
Analyzing Dalitz plots with different phases needs much more efforts on the experimental side -- and 
the theoretical side even more. One of us sees this obvious challenge as a job opportunity for young theorists, 
since most of the tools exist from Middle Energy Physics -- like dispersion relations -- it `just' needs more work.
\end{thebibliography}
\end{document}